\documentclass[fleqn,usenatbib]{mnras}

\usepackage[T1]{fontenc}
\DeclareRobustCommand{\VAN}[3]{#2}
\let\VANthebibliography\thebibliography
\def\thebibliography{\DeclareRobustCommand{\VAN}[3]{##3}\VANthebibliography}

\usepackage[dvipdfmx]{graphicx}	% Including figure files
\usepackage{amsmath}	% Advanced maths commands
\usepackage{bm}  
\usepackage{amssymb}	% Extra maths symbols
\usepackage{color}
\usepackage{ulem}

\graphicspath{{./figs/}}

\newcommand{\Msun}{{\rm  M_{\odot}}}

\newcommand{\Zsun}{Z_{\odot}}
\newcommand{\fesc}{f_{\rm esc}}

\newcommand{\Lya}{$\rm{Ly} \alpha$ }

\newcommand{\Mh}{M_{\rm h}}
\newcommand{\dmbh}{\dot{m}_{\rm BH}}
\newcommand{\Mbh}{{\rm M_{\rm BH}}}

\newcommand{\Msunyr}{{\rm \Msun~{\rm yr^{-1}}}}
\newcommand{\lya}{\rm Ly\alpha}
\newcommand{\Mstar}{{\rm M_{star}}}
\newcommand{\fgas}{{f_{\rm gas}}}
\newcommand{\Ssub}{S_{\rm 1.1mm}}
\newcommand{\Lir}{L_{\rm IR}}

\title[Forever22] 
{FOREVER22: galaxy formation in protocluster regions}
\author[Yajima et al. ]{Hidenobu Yajima$^{1}$\thanks{E-mail: yajima@ccs.tsukuba.ac.jp}, Makito Abe$^{1}$,  Sadegh Khochfar$^{2}$, Kentaro Nagamine$^{3,4,5}$, 
\newauthor Akio K. Inoue$^{6}$, Tadayuki Kodama$^{7}$, Shohei Arata$^{3}$, Claudio Dalla Vecchia$^{8}$,
\newauthor  Hajime Fukushima$^{1}$, Takuya Hashimoto$^{10}$, Nobunari Kashikawa$^{11}$,    
\newauthor  Mariko Kubo$^{12}$, Yuexing Li$^{13}$, Yuichi Matsuda$^{14, 15}$, Ken Mawatari$^{16}$, 
\newauthor Masami Ouchi$^{4, 14, 16}$, Hideki Umehata$^{17, 18}$ \\
$^{1}$Center for Computational Sciences, University of Tsukuba, Ten-nodai, 1-1-1 Tsukuba, Ibaraki 305-8577, Japan\\
$^{2}$Institute for Astronomy, University of Edinburgh, Royal Observatory, Edinburgh, EH9 3HJ, UK\\
$^{3}$ Department of Earth and Space Science, Graduate School of Science, Osaka University, Toyonaka, Osaka 560-0043, Japan\\
$^{4}$ Kavli IPMU (WPI), The University of Tokyo, 5-1-5 Kashiwanoha, Kashiwa, Chiba, 277-8583, Japan \\
$^{5}$ Department of Physics \& Astronomy, University of Nevada, Las Vegas, 4505 S. Maryland Pkwy, Las Vegas, NV 89154-4002, USA \\
$^{6}$Waseda Research Institute for Science and Engineering, Faculty of Science and Engineering, Waseda University, 3-4-1, \\
Okubo, Shinjuku, Tokyo 169-8555,
Japan\\
$^{7}$Astronomical Institute, Tohoku University, Sendai 980-8578, Japan \\
$^{8}$Instituto de Astrof\`{\i}sica de Canarias, C/V\`{\i}a L\`{a}ctea s/n, 38205 La Laguna, Tenerife, Spain\\
$^{9}$ Departamento de Astrof\'isica, Universidad de La Laguna, Av.~del Astrof\'isico Francisco S\'anchez s/n, 38206 La Laguna, Spain \\
$^{10}$ Tomonaga Center for the History of the Universe (TCHoU), Faculty of Pure and Applied Sciences, University of Tsukuba, \\
Tsukuba, Ibaraki 305-8571, Japan\\
$^{11}$Department of Astronomy, Graduate School of Science, The University of Tokyo, 7-3-1 Hongo, Bunkyo, Tokyo 113-0033, Japan\\
$^{12}$Research Center for Space and Cosmic Evolution, Ehime University, Matsuyama, Ehime 790-8577, Japan\\
$^{13}$Department of Astronomy and Astrophysics, The Pennsylvania State University, University Park, PA 16802, USA\\
$^{14}$National Astronomical Observatory of Japan, 2-21-1 Osawa, Mitaka, Tokyo 181-8588, Japan\\
$^{15}$Department of Astronomical Science, SOKENDAI (Graduate University for Advanced Studies), Osawa 2-21-1, Mitaka, Tokyo181-8588, Japan\\
$^{16}$Institute for Cosmic Ray Research, The University of Tokyo, 5-1-5 Kashiwa-no-Ha, Kashiwa, Chiba 277-8582, Japan\\
$^{17}$RIKEN Cluster for Pioneering Research, 2-1 Hirosawa, Wako-shi, Saitama 351-0198, Japan\\
$^{18}$Institute of Astronomy, School of Science, The University of Tokyo, 2-21-1 Osawa, Mitaka, Tokyo 181-0015, Japan
}

\begin{document}

\date{Accepted XXX. Received YYY; in original form ZZZ}

\pagerange{\pageref{firstpage}--\pageref{lastpage}} \pubyear{2022}

\maketitle

\label{firstpage}

%----------------------------------------------------------------------
%
% Abstract
%
%----------------------------------------------------------------------
\begin{abstract}
We present results from a new cosmological hydrodynamics simulation campaign of protocluster (PC) regions, 
FOREVER22: FORmation and EVolution of galaxies in Extremely-overdense Regions motivated by SSA22. 
The simulations cover a wide range of cosmological scales 
using three different zoom set-ups in a parent volume of $(714.2~\rm cMpc)^{3}$:
PCR (Proto-Cluster Region;  $V= (28.6~{\rm cMpc})^{3} $, SPH particle mass, $m_{\rm{SPH}} = 4.1 \times 10^{6}~\Msun$ and final redshift, $z_{\rm end}=2.0$), BCG (Brightest proto-Cluster Galaxy;  $V \sim (10~{\rm cMpc})^{3} $, $m_{\rm SPH} = 5.0\times10^{5}~\Msun$ and $z_{\rm end}=4.0$ ), and First ( $V \sim (3~{\rm cMpc})^{3} $, $m_{\rm SPH} = 7.9 \times 10^{3}~\Msun$ and $z_{\rm end}=9.5$) runs, that allow to focus on different aspects of galaxy formation.
In the PCR runs, we follow 10 PCs, each harbouring 1 - 4 SMBHs with $\Mbh \ge 10^{9}~\Msun$.
One of the PC cores shows a spatially close arrangement of seven starburst galaxies with ${\rm SFR} \gtrsim 100~\Msunyr$ each, that are dust-obscured and would appear as submillimeter galaxies with flux $\gtrsim 1~$ mJy at $1.1~ \rm mm$ in observations. The BCG runs show that the total SFRs of haloes hosting BCGs are affected by AGN feedback, but exceed $1000~\Msunyr$ at $z \lesssim 6$.
The First runs resolve mini-haloes hosting population (Pop) III stars and we show that, in  PC regions, the dominant stellar population changes from Pop III to Pop II at $z \gtrsim 20$, and the first galaxies with ${\rm SFR} \gtrsim 18~\Msunyr$ form at $z \sim 10$. These can be prime targets for future observations with the {\it James Webb Space Telescope}.
Our simulations successfully reproduce  the global star formation activities in observed PCs and suggest that PCs can  kickstart cosmic reionization.
\end{abstract}
%----------------------------------------------------------------------
%
% Keywords
%
%----------------------------------------------------------------------

%----------------------------------------------------------------------
%
% Section 1: Introduction
%
%----------------------------------------------------------------------
\section{Introduction}
Understanding galaxy evolution in the early Universe 
is one of the major goals in astrophysics.  
The recent development of observational facilities has allowed us to probe high-redshift galaxies and the large-scale structure of the Universe. 
Observations with optical/near-infrared telescopes have successfully observed numerous high-redshift galaxies with the drop-out technique, called ``Lyman-break galaxies (LBGs)" \citep[e.g.,][]{Shapley11, Bouwens15, Oesch16, Ouchi18}. 
Also, some of them have been detected with strong \Lya or H$\alpha$ lines originated from ionized gas due to young stars, called ``Lyman-alpha emitters (LAEs)" and ``H {$\rm \alpha$} emitters (HAEs)" \citep[e.g.,][]{Iye06, Finkelstein13, Ono18, Hayashi20}. 
This progress has led to a consensus on the evolution of the 
cosmic star formation rate densities (CSFRDs) between $z = 0$ and $z  \sim 10$ 
\citep[e.g.,][]{Madau14, Bouwens15, Bouwens20}, although there is still some uncertainty at $z \gtrsim 5$ \citep{Khusanova20, Talia21}.  
In addition to the detections of direct stellar radiation, recent observations using submillimeter telescopes, e.g.,  the Atacama Large Millimeter/submillimeter Array (ALMA) have detected dust thermal emission from distant galaxies, called ``submillimeter galaxies (SMGs)"  
\citep[e.g.,][]{Chapman05, Riechers13, Hatsukade18, Marrone18}. As the galaxy mass increases, star-forming regions can be enshrouded 
by dust because of higher metallicity and dust content \citep{Casey14}. 
Therefore the submillimeter flux can be a powerful tool to probe massive galaxies with active star formation. 
Moreover, [OIII] 88 $\micron$ or [CII] 158 $\micron$ lines from distant galaxies have been successfully detected with ALMA \citep{Capak15, Inoue16}. E.g. \citet{Hashimoto18} have spectroscopically confirmed the most distant galaxies at $z = 9.1$ via the detection of the  [OIII] 88 $\micron$ line \citep[see also the most distant galaxy without  line detection at $z=11.1$:][]{Oesch16}. 
While it seems evident that the  various observational properties are likely linked with fundamental physical properties such as star formation, distribution of gas and dust and gas kinematics, the exact connection is still poorly understood.  

According to the current standard paradigm of structure formation, galaxies evolve via mergers and matter accretion from large-scale filaments \citep[e.g.,][]{Springel06}. The growth rates of galaxies sensitively depend on formation sites. In overdense regions, galaxies rapidly grow, while galaxies in void regions do slowly \citep[e.g.,][]{Benson03}. 
Therefore, understanding the environmental effects can be a key to reveal various evolutionary scenarios for  high-redshift galaxies.  
%--Protocluster----\\
In overdense regions, galaxies cluster on shorter length scales 
\citep[see the review by][]{Overzier16}.
Theoretical models based on cosmological $N$-body simulations indicated that regions with a higher level of clustering at early times will evolve into present-day   
galaxy clusters at $z \sim 0$ \citep{Chiang17}. 
It therefore makes sense to associate such regions as  
``protoclusters (PCs)" regions, a term we will use throughout this paper. 
As a unique massive protocluster in the early Universe, the region SSA22 
at $z = 3.1$ has been investigated by various observational techniques. 
E.g. the large scale filamentary structure around SSA22 has been studied   using the spatial distributions of LAEs 
\citep{Hayashino04, Matsuda04}. 
Extended Ly$\alpha$ sources, called ``Lyman-alpha blobs (LABs)" , 
 with sizes of $\gtrsim 100~\rm kpc$ have been reported at the core of SSA22 \citep{Steidel00, Matsuda12}. 
\citet{Tamura09} showed that SMGs distributed near LAEs in SSA22 \citep[see also,][]{Umehata15, Umehata18, Umehata19}. 
These observations suggest that various galaxies can form and coexist in such an overdense regions. 
Therefore, PCs have the potential to be laboratories to understand the diversity of  galaxy evolution. 

A recent wide survey with Subaru Hyper Supreme Cam observed 
$\gtrsim 200$ candidates of protoclusters composed of LBGs at $z \gtrsim 4$ \citep{Toshikawa18}. 
\citet{Harikane19} discovered a protocluster with LAEs/LABs at $z > 6$ \citep[see also,][]{Ishigaki16}. 
Also, \citet{Miller18} discovered a clustered region of dusty starburst galaxies at $z = 4$
where the total star formation rate of observed galaxies at the PC core exceeded $6000~\Msunyr$ \citep[see also,][]{Oteo18}. 
Thus, recent observations have allowed us to study galaxy formation in PCs, and as such the onset of environmental effects. 

Combining cosmological $N$-body simulations and semi-analytical galaxy formation models, \citet{Chiang17} investigated star formation in PCs. 
They suggested that PCs contributed significantly 
to the cosmic star formation rate density (CSFRD) at high-redshift $z \gtrsim 2$ and trigger cosmic reionization. 
Recent hydrodynamics simulations of galaxy formation in large-scale structures have studied star formation, gas dynamics, and stellar/AGN feedback processes \citep[e.g.,][]{Vogelsberger14}. 
Using the moving mesh hydrodynamics code {\sc arepo} \citep{Springel10}, the Illustris/illustrisTNG projects showed results for cosmological hydrodynamics simulations of cosmic volumes of $(\rm 50 - 100~cMpc)^{3}$, and successfully reproduced various properties  of the local galaxies population \citep[e.g.,][]{Nelson18, Pillepich18}.
 In a similar project ({\sc Eagle} project) using a smoothed particle hydrodynamics (SPH) code \citet[][ hereafter S15]{Schaye15} 
reproduced physical properties of the local galaxy population.  
They introduced sub-grid models associated with star formation and black holes and their feedback processes with tuned parameters \citep[see also,][]{Crain15}. As indicated by the comparison between stellar and halo mass functions, 
stellar feedback can regulate star formation in low-mass galaxies and the feedback from active galactic nuclei (AGNs) regulates star formation in massive galaxies. The above projects applied feedback models to successfully regulate the star formation activities of low-mass and massive galaxies  appropriately. 
\citet{Dubois16} studied the impacts of AGNs on galaxies and the circum-galactic medium (CGM) in the {\sc Horizon-AGN} simulation with the adaptive mesh refinement code, {\sc Ramses} \citep{Teyssier02}.  
They indicated that the morphology of massive galaxies sensitively depended on  AGN feedback \citep[see also,][]{Sijacki15, DiMatteo17,Tremmel17}.
Thus, the recent developments of simulation codes and sub-grid models have allowed us to model galaxies reproducing statistical properties of  local galaxies as inferred from an average over different environments  and study physical processes to determine star formation and BH activities and the distribution of gas and stars. 
 
On the other hand,  galaxy evolution in overdense regions has not been studied and understood well. 
\citet{Barnes17} consider a huge cosmological volume of $(3.2~\rm Gpc)^{3}$ and selected 30 galaxy clusters at $z=0$. 
They studied the 30 clusters with zoom-in simulations with the calculation code developed in EAGLE project, which is called the Cluster-EAGLE (C-EAGLE) project \citep[see also,][]{Bahe17}. Their simulations reproduced stellar and BH components in local galaxy clusters, while the gas fraction was too high. Also, 
\citet{Cui18} investigated statistical properties of galaxy clusters for a sample of 324 clusters (The Three Hundred project) 
based on zoom-in simulations with a modified version of SPH code, {\sc gadget2} \citep{Springel05e}. 
They showed the baryonic fraction of the clusters matched  observations, 
while there were some differences in the masses of member galaxies and their colors. 
Thus, while recent simulations successfully reproduced observed properties of local galaxy clusters partially, 
baryonic physics in the overdense regions is still puzzling.  
Recently, \citet{Trebitsch20} studied a protocluster, the most massive halo in a volume of $(100~\rm cMpc)^{3}$, 
and investigated the contribution to cosmic reionization in their simulation, the {\sc obelisk}, which is the updated version of the  {\sc horizon-agn} project. 
They investigated the onset of cosmic reionization in an overdense region and showed that  hydrogen reionization was completed by galaxies, only at $z \sim 4$ radiation from black holes started to play an important role.  

Here, we introduce a new simulation project, {\sc forever22} (FORmation and EVolution of galaxies in Extremerly-overdense Regions motivated by SSA22). 
In this project, we study galaxy evolution in protoclusters and the formation mechanism of observed galaxies, LAEs, LBGs, SMG, passive galaxies, and QSOs in the protoclusters at redhsifts $ z \ge 2$. Using a large volume of $(714~\rm cMpc)^{3}$, we select the top 10 massive haloes and study the statistical  properties of galaxies in them, baryonic physics and radiative properties. 
The {\sc forever22} consists of three simulation sets with  different resolutions and volumes: PCR (Proto-Cluster Region), BCG (Brightest proto-Cluster Galaxy), and First runs. 
By using these three series, we can investigate both the statistical nature and the small scale baryonic physics with stellar/AGN feedback. 

Besides, we carry out multi-wavelength radiative transfer simulations that can 
calculate the properties of continuum flux from X-ray to radio, Lyman continuum, 
\Lya, [O{\sc iii}], [C{\sc ii}], CO lines. Thus, we can directly compare the simulations with recent observations 
with optical/NIR telescopes (e.g., Subaru, Keck, Hubble Space Telescopes) and radio telescopes (e.g., ALMA), 
and also predict for future missions (e.g., James Webb Space Telescope). 

Our paper is organized as follows. In Section 2, we describe the numerical methods and setup of our simulations.  We present our simulation results of the PCR runs in Section 3.1 - 3.4. 
The results of the BCG and First runs are shown in Section 3.4 and 3.5, respectively. 
In Section 4, we discuss our results and summarize our main conclusions.

%----------------------------------------------------------------------
%
% Section 2:  Model and Method
%
%----------------------------------------------------------------------

%%%-----------------------
\begin{table*}
\begin{center}
\begin{tabular}{ccccccccc}
\hline
Halo ID &  $\Mh \;[\Msun/h]$ at $z_{\rm end}~(z=0, 3)$& $m_{\rm gas}\;[\Msun/h]$ & $m_{\rm DM}\;[\Msun/h]$ & $\epsilon_{\rm min}\:[{\rm~kpc}/h]$   & $z_{\rm end}$ \\
\hline
PCR0      &$1.9 \times10^{14} ~ (1.4 \times 10^{15}, ~8.1\times10^{13})$   & $2.9\times10^{6}$ & $1.6\times10^{7}$ & 2.0  &2 \\
PCR1      &$1.5 \times10^{14}~(1.2 \times 10^{15}, ~5.9 \times 10^{13})$   & $2.9\times10^{6}$ & $1.6\times10^{7}$ & 2.0  &2 \\
PCR2   & $1.2\times10^{14}~(5.2 \times 10^{14}, ~5.6\times 10^{13})$ & $2.9\times10^{6}$  & $1.6 \times10^{7}$ & 2.0   &2 \\
PCR3  & $1.2\times10^{14}~(8.1 \times 10^{14}, ~ 2.0 \times 10^{13})$ & $2.9\times10^{6}$  & $1.6 \times10^{7}$ & 2.0   &2 \\
PCR4      & $1.1\times10^{14}~(1.1\times 10^{15}, ~ \times 10^{13})$   & $2.9\times10^{6}$ & $1.6\times10^{7}$ & 2.0  &2  \\
PCR5      & $1.0\times10^{14}~(6.7 \times 10^{14}, ~3.5 \times 10^{13})$   & $2.9\times10^{6}$ & $1.6\times10^{7}$ & 2.0 &2  \\
PCR6      & $9.9\times10^{13}~(5.9 \times 10^{14}, ~4.6 \times 10^{13})$   & $2.9\times10^{6}$ & $1.6\times10^{7}$ & 2.0 &2  \\
PCR7      & $9.6\times10^{13}~(6.3 \times 10^{14}, ~3.3 \times 10^{13})$   & $2.9\times10^{6}$ & $1.6\times10^{7}$ & 2.0 &2  \\
PCR8      & $9.1\times10^{13}~(5.6 \times 10^{14}, ~5.1 \times 10^{13})$   & $2.9\times10^{6}$ & $1.6\times10^{7}$ & 2.0 &2  \\
PCR9      & $9.1\times10^{13}~(1.2 \times 10^{15}, ~2.0 \times 10^{13})$   & $2.9\times10^{6}$ & $1.6\times10^{7}$ & 2.0 &2  \\
MF   & $1.4 \times10^{13}~(6.1 \times 10^{12})$  & $2.9\times10^{6}$  & $1.6\times10^{7}$ & 2.0  &2 \\
BCG0  & $2.0 \times10^{13}$  & $3.5\times10^{5}$  & $2.0\times10^{6}$ & 1.0  &4 \\
BCG1  & $2.9 \times10^{13}$  & $3.5\times10^{5}$  & $2.0\times10^{6}$ & 1.0  &4 \\
BCG2  & $2.8 \times10^{13}$  & $3.5\times10^{5}$  & $2.0\times10^{6}$ & 1.0  &4 \\
BCG3  & $5.4 \times10^{12}$  & $3.5\times10^{5}$  & $2.0\times10^{6}$ & 1.0  &4 \\
BCG4  & $2.0 \times10^{13}$  & $3.5\times10^{5}$  & $2.0\times10^{6}$ & 1.0  &4 \\
BCG5  & $1.8 \times10^{13}$  & $3.5\times10^{5}$  & $2.0\times10^{6}$ & 1.0  &4 \\
BCG6  & $1.7 \times10^{13}$  & $3.5\times10^{5}$  & $2.0\times10^{6}$ & 1.0  &4 \\
BCG7  & $9.1 \times10^{12}$  & $3.5\times10^{5}$  & $2.0\times10^{6}$ & 1.0  &4 \\
BCG8  & $1.5 \times10^{13}$  & $3.5\times10^{5}$  & $2.0\times10^{6}$ & 1.0  &4 \\
BCG9  & $6.0 \times10^{12}$  & $3.5\times10^{5}$  & $2.0\times10^{6}$ & 1.0  &4 \\
BCG0noAGN      &$2.0 \times10^{13}$   & $3.5\times10^{5}$ & $2.0\times10^{6}$ & 1.0  &4\\
BCG0spEdd      &$2.0 \times10^{13}$   & $3.5\times10^{5}$ & $2.0\times10^{6}$ & 1.0  &4 \\
First0  & $3.5 \times10^{11}$  & $5.5\times10^{3}$  & $3.1 \times10^{4}$ & 0.2  &9.5 \\
First1  & $3.2 \times10^{11}$  & $5.5 \times10^{3}$  & $3.1\times10^{4}$ & 0.2  &9.5 \\
\hline
\end{tabular}
\caption{
Parameters of zoom-in cosmological hydrodynamic simulations: 
(1) $\Mh$ is the halo mass at the final redshift ($z_{\rm end}$). (2) $m_{\rm gas}$ is the initial mass of gas particles. (3) $m_{\rm DM}$ is the dark matter particle mass. 
(4) $\epsilon_{\rm min}$ is the gravitational softening length in comoving units.
}
\label{table:halo}
\end{center}
\end{table*}
%%%-----------------------

%\section{Methodology}
\section{The FOREVER22 simulation}
\label{sec:f22}

We utilize the SPH code {\sc gadget-3} \citep{Springel05e}
with the modifications developed in the {\it Overwhelmingly Large Simulations} (OWLS) project \citep{Schaye10}.
Following the EAGLE project \citep{Schaye15}, we update the star formation and supernova (SN) feedback models. 
This code was also modified to handle the formation of population III (Pop III) stars, Lyman-Werner feedback and non-equilibrium primordial chemistry  in the {\it First Billion Year} (FiBY) project \citep{Johnson13, Paardekooper15}. 
Metal line cooling is also considered based on the equilibrium state with the UV background \citep{Wiersma09}.
The new models as part of the FiBY project have been used to e.g. investigate the cosmic star-formation rate density of Pop III and Pop II stars, Lyman continuum leakage from high-redshift dwarf galaxies, dust extinction in high redshift galaxies, globular cluster formation and   statistical properties of direct-collapse black holes \citep[e.g.,][]{Paardekooper13, Agarwal15, Elliot15, Cullen17, Phipps20}. 
In this project, we  add models to calculate the radiative feedback from young stars and kinetic feedback from massive black holes and the growth/destruction of dust grains. 

The FOREVER22 project consists of three series of simulations: PCR, BCG, and First runs. 
Recent observational wide surveys reveal the large-scale structures around protoclusters \citep[e.g.,][]{Kikuta19}, while at the same time  high-angular resolution  ALMA observations can resolve giant gas clumps or spiral arms in high-redshift galaxies \citep[e.g.,][]{Tadaki18}. 
State-of-the-art simulations still struggle resolving both large-scale structures and small-scale gas clumps simultaneously. To overcome this limitation  we designed the above mentioned runs to investigate the statistical properties of galaxies in protoclusters and their detailed structure evolution and feedback. 
The resolutions and parameters are summarised in Table~\ref{table:halo} and more details on the individual runs will be given in the following sections. \\ 

\noindent
{\it   $\bullet$ Proto-Cluster Region  (PCR) runs }\\ 
In the PCR runs, we consider $(28.6~\rm cMpc)^{3}$ volumes to investigate the statistical nature of galaxies in  protoclusters and the large-scale structures around them, where $\rm cMpc$ is comoving Mpc. 
We use the {\sc music} code \citep{Hahn11} to create the initial conditions in this work. 
The volume of the entire calculation box is $(714.2~\rm cMpc)^{3}$. 
First, we carry out a $N$-body simulation in the entire box with $256^3$ dark matter particles.
Then, we choose the top 10 most massive haloes in the box at $z=2$ and make zoom-in initial conditions with a side length of $28.6~\rm cMpc$. All simulations start from $z=100$.
To highlight the effect of the environment,
%we also choose three mean density regions and take the mean of them (MF run). 
we also choose three mean density regions with the same volume and resolution and derive statistical properties of galaxies. 
The mean density regions are randomly selected from the entire simulation box. We confirm that these regions are not overlapping with each other and their matter densities are close to the cosmic mean.
We carry out hydrodynamics simulations down to $z=2$ and study the statistical nature of galaxies  and compare them with those in the mean density run.  
Note that, in the regions near the boundaries between the zoom-in  and the outside low-resolution regions, hydrodynamic and gravitational forces might not be accurate. Therefore, we use galaxies only in the inner regions with a volume of $(25.7~\rm cMpc)^{3}$ to analyze statistical properties, like e.g. the stellar mass function. We confirm massive particles originally in the outside low-resolution regions do not enter the inner regions. 

For all simulations, a total of 200 snapshots are output with the same time interval from $z=100$ to the final redshift. For example, in the case of PC runs, the time interval is $\Delta t \sim 16~\rm Myr$. 
According to \citet{Chiang17}, matter in a volume with radius of $\sim 10~\rm cMpc$ is typically incorporated into a galaxy cluster at $z \sim 0$. Our zoom-in regions cover such  volumes.
Note that, however, the volume depends on the cluster mass at $z=0$ \citep{Muldrew15, Lovell18}. In the cases of massive galaxy clusters ($\gtrsim 10^{15}~\Msun$), the region enclosing all matter is larger than $10~\rm cMpc$.  
To estimate the masses of galaxy clusters in the PCRs, we perform $N$-body simulations down to $z=0$ with the initial zoom-in volume of $(57.1~\rm cMpc)^{3}$ and show the values in Table~\ref{table:halo}. 
We find all PCRs form  massive haloes of $> 10^{14}~\Msun$ and the halo mass exceeds $10^{15}~\Msun$ in the cases of PCR0, PCR1, PCR4 and PCR9. 
Then, we estimate radii as a function of redshift within which 90 percent of the matter in the galaxy clusters at $z=0$ is enclosed. We find that the radii range $\sim 15 - 25~\rm cMpc$ which is somewhat larger than the size of the initial zoom-in regions in the PCR runs. Therefore, the PCRs  do not cover all building blocks of the galaxy clusters at $z=0$. 
The zoom-in regions enclose $\sim 46 - 80$ percent of the mass content of the descendant galaxy clusters at $z=0$ (e.g., 46 percent for PCR0, 62 percent for PCR3, and 80 percent for PCR2).
However, the zoom-in regions contain the main progenitors and other massive galaxies that form the main building blocks. 
Since we aim at revealing the evolution of massive galaxies in the overdense regions in this project, the size of zoom-in volume is not problematic. In addition, we confirm that the masses of the most massive haloes in the PCRs at $z=2$ are the same as  in the $N$-body simulations. 
In Section~\ref{sec:BCG}, we also compare redshift evolutions of SFR, stellar and BH masses between PCR and BCG runs of which the resolution and the zoom-in volume are different.

\begin{figure*}
	\begin{center}
		\includegraphics[width=18cm]{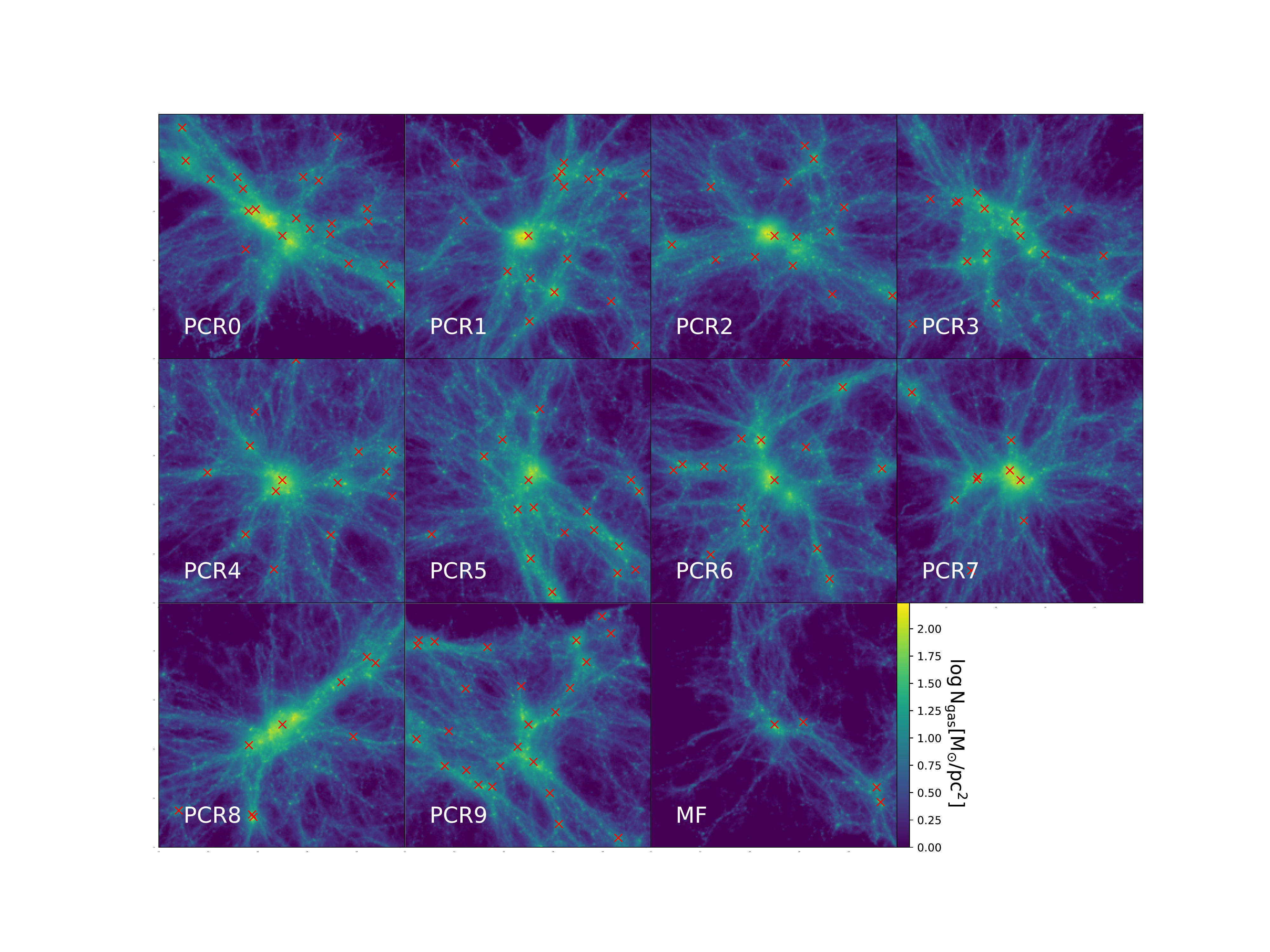}
	\end{center}
	\caption{		
	Gas structures in PCR and MF runs at $z=3$. The color represents gas column density with thickness of 10 cMpc. 
	The box size is $\rm 10~cMpc ~\times 10~cMpc$. Red crosses indicate the centre of mass of massive haloes with $\Mh \ge 10^{12}~\Msun$. 
		 }
	\label{fig:pcrmap}
\end{figure*}

Figure~\ref{fig:pcrmap} shows the gas distribution and positions of haloes with mass greater than $10^{12}~\Msun$. The centre of each panel is set at the centre of mass of all massive haloes with $\Mh \ge 10^{12}~\Msun$.
We find massive haloes form in nodes or cross points of large-scale filaments. The large-scale structures 
show variations. For example, PCR0 and PCR8 have a few thick filaments and the shapes look elongated. On the other hand, PCR4 and PCR7 consist of many thin filaments, and the shapes are isotropic. Unlike PCR regions, MF has only four massive haloes and no pronounced filaments are seen. 
%In the case of major mergers between haloes, the  peaks of the gas column density can be somewhat shifted from the centres of mass of haloes.
In some cases, the  peaks of the gas column density are somewhat shifted from the centres of mass of haloes. In these cases, multiple massive stellar components are distributed in a halo. Therefore, these haloes are likely undergoing a merger.
\\

\noindent
{\it  $\bullet$ Brightest proto-Cluster Galaxy (BCG) runs}\\
To study gas dynamics in massive galaxies, we increase the mass resolution while at the same time the zoom-in region is limited to cover the most massive haloes only in each PCR region. 
The masses of gas and dark matter particles are $3.5 \times 10^{5}$ and $2.0 \times 10^{6}~h^{-1}\;\Msun$ respectively, 
which are 8 times lower than in the PCR runs. We follow the evolution of the haloes down to $z=4$ when they become massive with the stellar and black hole masses of $\gtrsim 10^{11}$ and $\gtrsim 10^{8}~\Msun$. 
The zoom-in initial conditions are constructed ensuring they enclose all dark matter particles in a halo at $z=4$. The sizes of the initial zoom-in regions are $\sim 7-14~\rm cMpc$. We confirm that there are no differences in the halo masses at $z=4$ between BCG and PCR runs. 
We also use the  BCG runs to investigate the impact of stellar and AGN feedback on galaxy evolution. 
\\

\noindent 
{\it $\bullet$ First galaxy (First) runs} \\
The First runs are composed of  two zoom-in simulations focusing on the formation of the first galaxies  at $z \ge 9$.  
By making zoom-in initial conditions covering the most massive haloes in PCR0 and PCR1 regions at $z=9.5$, we increase the mass resolutions of gas and DM to $5.5 \times 10^{3}$ and $3.1 \times 10^{4}~h^{-1}\;\Msun$, which can resolve mini-haloes hosting Pop III stars. The sizes of the initial zoom-in regions are $\sim 3~\rm cMpc$. 
In the First runs in contrast to the other runs we also consider non-equilibrium chemistry of primordial gas to follow the gas collapse in the mini-haloes via $\rm H_{2}$ cooling. 

The FiBY project, \citet{Johnson13} presented  the cosmic star formation rate densities of Pop III and Pop II stars in a mean density environment. Due to the metal enrichment, Pop II star formation becomes dominant at $z \lesssim 10$. 
In our runs the transition from Pop III to Pop II stars likely occurs earlier than in the mean-density FiBY runs due to rapid metal enrichment via type-II supernovae (SNe). Using the First runs, we study the metal enrichment in protocluster regions and formation of first galaxies with Pop II stars. Also, upcoming telescopes, e.g., James Webb Space Telescope (JWST) aim at detecting the first galaxies at $z \gtrsim 10$. 
Since galaxies in the over-dense regions are likely to have a high star formation rate (SFR), they can be plausible candidates for future observations. Therefore we investigate the brightness and observability of the first galaxies in PCs. \\

We use a friend-of-friend (FOF) group finder to identify haloes on-the-fly.  
In massive haloes, there are some satellite galaxies. We utilize {\sc subfind} \citep{Springel05e} to identify member galaxies in haloes in post-processing. 
We adopt following cosmological parameters that are consistent with the current cosmic microwave background observations: $\Omega_{\rm M}=0.3$, $\Omega_{\rm b}=0.045$, $\Omega_{\rm \Lambda}=0.7$, $n_{\rm s}=0.965$, $\sigma_{8}=0.82$, and $h=0.7$ \citep{Komatsu11, Planck16, Planck20}. 

%%%-----------------------
%%%-----------------------

\subsection{Star Formation}

We follow a star formation (SF) model developed in \citet{Schaye08} which was used in OWLS and EAGLE projects. 
This SF model is based on the Kennicutt-Schmidt law of local galaxies, 
i.e., SFR surface density is proportional to gas surface density. 
\citet{Schaye08} assume the disk scale height to be equal to the Jeans length and model the local SFR based on the local ISM pressure:
\begin{equation}
\dot{m}_{*} = m_{\rm g}A 
\left( 1~\rm \Msun~pc^{-2} \right)^{-n} 
\left(  \frac{\gamma}{G} f_{\rm g} P \right)^{(n-1)/2},
\end{equation}
where $m_{\rm g}$ is the mass of the gas particle, $\gamma= 5/3$ is the ratio of specific heats,
$f_{\rm g}$ is the gas mass fraction in the self-gravitating galactic disc,  and $P$ is the total ISM pressure. 
The free parameters in this SF model are the amplitude $A$ and the power-law index $n$.
These parameters are related to the Kennicutt-Schmidt law, 
\begin{equation}
\dot{\Sigma}_{*} = A \left(\frac{\Sigma_{\rm gas} }{ 1 \Msun~{\rm pc^{-2}}} \right)^{n}. 
\end{equation}
Local normal star-forming galaxies follow $A_{\rm local}=2.5\times10^{-4}~\rm \Msun~yr^{-1}~kpc^{-2}$
and $n=1.4$ for a Salpeter IMF \citep{Kennicutt98b}. 
Note that, the amplitude should be changed by a factor $1/1.65$ in the case of the Chabrier IMF, i.e., $A_{\rm local,Chab}=1.5 \times10^{-4}~\rm \Msun~yr^{-1}~kpc^{-2}$. 
In this work, we use the Chabrier IMF with a mass range of $0.1 - 100~\rm \Msun$.
\citet{Schaye10} reproduced the observed cosmic star formation rate density (SFRD) using cosmological SPH simulations with this SF model and the parameters $A=1.5 \times 10^{-4}~\rm \Msun~yr^{-1}~kpc^{-2}$ and $n=1.4$ \citep[see also,][]{Schaye15}.  
As in EAGLE, we change the slope $n$ to $2.0$ for high-density gas with $n_{\rm H} > 10^{3}~\rm cm^{-3}$ and 
use the threshold density depending on local metallicity as $n_{\rm H} = n_{0}~{\rm cm^{-3}}\left( \frac{Z}{0.002}\right)^{-0.64}$
where we set $n_{0}=0.1$ for PCR and BCG runs and $10.0$ for First runs. 

For gas at densities  $n_{\rm H} > n_{0}$, we use an effective equation of state with an effective adiabatic index $\gamma_{\rm eff}=4/3$. The floor temperatures at $n_{\rm H} = n_{0}$ are $8000~\rm K$ for PCR and BCG runs and $1000~\rm K$ for First runs. 

In BCG and First runs, we follow Pop III star formation. If the gas phase metallicity of star forming gas  is lower than $1.5\times 10^{-4}~\Zsun$ \citep{Bromm03, Omukai05}, Pop III stars form with an initial mass function (IMF) $dn \propto M^{-2.35} dM$  within the mass range $10- 500~\Msun$.  Due to the higher typical stellar mass,  Pop III stars give strong feedback to the surrounding gas and induce metal enrichment rapidly.
Since we set a minimum mass of $10~\Msun$, all Pop III stars will  end as SNe or direct collapse BHs. Therefore, we do not consider  energetic feedback and metal pollution via Type Ia SNe and the AGB phase from Pop III stars.
Some Pop III stars are likely to be single stellar-mass BHs or high-mass X-ray binaries (HMXBs) at the end of  their lifetime and will suppress star formation in the first galaxies \citep[e.g.,][]{Jeon14}. In this work, we do not take  remnant BHs and HMXBs into account. 

\subsection{UV background radiation}

As the cosmic star formation rate density (CSFRD) increases, the  universe is filled with UV background (UVB) radiation \citep{Haardt96, Haardt01, Faucher09}. The UVB heats the inter-galactic medium (IGM) and changes the ionization states of primordial gas and metals.  
The cooling rate is estimated from the assumption of equilibrium (collisional or photoionization) for each metal species. 
The metal-line cooling is considered for each metal species using a pre-calculated table by {\sc cloudy} v07.02 code \citep{Ferland00}.  
At $z \lesssim 10$, galaxies are irradiated by the UVB, and it penetrates into the gas with $n_{\rm H} < 0.01 ~\rm cm^{-3}$ which is the threshold density for self-shielding as derived by \citet{Nagamine10a} and \citet{Yajima12h} based on the radiative transfer calculations of the UVB.  
We switch from collisional to photoionization equilibrium cooling tables once the UVB ionizes the gas \citep[see][for details]{Johnson13}. We use the UVB of \citet{Haardt01} in our simulations.
The clustering of galaxies and the high star formation activity in the protocluster regions can boost the local UV radiation field. In this work, we do not take into account local fluctuations of the UVB.

\subsection{Stellar Radiation Feedback}
 
 \noindent
{\it   $\bullet$ Photo-ionization feedback }\\
We take account of feedback from stars by considering the photo-ionization heating and radiation pressure on dust. This radiative feedback mainly originates from young star clusters. Therefore, in this work, we take stellar particles with an age of $\le 10~\rm Myr$ into account as the sources of the radiative feedback.
 We assume a black-body spectrum with $T=10^{5}~\rm K$ for Pop III stars and a synthesized SED with a Chabrier IMF with zero age for Pop II stars, and estimate the photon production rate from a stellar particle. 
 Once the stellar age exceeds $10~\rm Myr$, we turn off the radiative feedback and then consider supernova feedback as explained below.
 
 The photo-ionization heats the gas to $\gtrsim 10^{4}~\rm K$, resulting in the expansions of H{\sc ii} bubbles due to the higher thermal pressure if the pressure of surrounding gas is lower. 
 We estimate the ionized region by solving the balance between the ionizing photon production rate ($\dot{N}_{\rm ion}$) from young star-clusters and the total recombination rate of the ionized gas as:
 \begin{equation}
 \dot{N}_{\rm ion} = \sum_{i=1}^{n} \alpha_{\rm B} n_{\rm HII}^{i} n_{\rm e}^{i} \frac{m_{\rm gas}^{i}}{\rho_{\rm gas}^{i}},
 \end{equation}
 where $\dot{N}_{\rm ion}$ is the photon production rate of a stellar particle, $\alpha_{\rm B}$ is the case-B recombination coefficient, 
 $n_{\rm HII}^{i}$ and $n_{\rm e}^{i}$ are the ionized hydrogen and electron number densities of $i$-th SPH particle. In the ionized region, we set $n_{\rm HII} = n_{\rm e} = n_{\rm H}$, where $n_{\rm H}$ is the total hydrogen number density. Here we evaluate the volume by the gas mass of $i$-th SPH particle $m_{\rm gas}^{i}$ and its mass density $\rho_{\rm gas}^{i}$. 
 We sum up the total recombination rate of the surrounding gas from the nearest gas particle ($i=1$) in turn. 
 If there are several stellar particles in a small area, the ionized regions can overlap each other and make larger ionized bubbles. In this work, we do not consider the overlap effect.
 The temperature of the ionized gas is set to $T_{\rm HII} = 3 \times 10^{4}~\rm K$. 
 The temperature of ionized regions can change between $\sim 1 \times 10^{4} < T_{\rm HII} < \sim 3 \times 10^{4}~\rm K$ depending on the stellar metallicity. We consider the case of very low-metallicity or Pop III stars.
 Note that, in the case of PCR and BCG, the pressure of high-density regions can be higher than the case considering $T_{\rm HII}$ because of the effective equation-of-state (EoS) model. Therefore, the photo-ionization heating works only for the low-density environments at $n_{\rm H}\lesssim 5 ~\rm cm^{-3}$.
 If the recombination rate of the nearest gas-particle alone is higher than $\dot{N}_{\rm ion}$, 
 only the nearest particle is recognized as in the ionized region. 
We prohibit the star formation in the ionized region. 
The photoionization model is used for all simulations. Note that, ionized regions around stellar particles are not resolved well in the cases of PCR and BCG runs. Thus, the suppression of star formation in the nearest gas particles is the main effect, while the thermal pressure in the ionized region works in the First runs. 
\\
 
\noindent
{\it   $\bullet$ Radiation pressure on dust}\\
 A part of UV radiation from young stars is absorbed by dust, which gives outward momentum to gas \citep[e.g.,][]{Murray05, Yajima17b}.  
 We here estimate the mean free path of UV continuum photons in dusty gas as
 $l_{\rm m.f.p} = \frac{1}{\kappa_{\rm d} \rho_{\rm gas}}$, 
 where $\kappa_{\rm d}$ is the absorption coefficiency. Here we set $\kappa_{\rm d}=2.5 \times 10^{2}~ \rm cm^{-2}~g^{-1} \left( Z / \Zsun \right)$, 
 which is corresponding to the silicate dust with the size of $\sim 0.1~\rm \mu m$
 and the dust-to-gas mass ratio of $\sim 0.01$ corresponding to solar abundance \citep{Yajima17b}.  
 Within $l_{\rm m.f.p}$, we can assume to be in the optically thin limit and estimate the radiation force as
 \begin{equation}
 {\bf F}_{\rm rad} =  \frac{\rho_{\rm gas} \kappa_{\rm d} L_{\rm UV}}{4 \pi r^{2} c} \frac{\bf r}{r}, 
 \end{equation}
where $L_{\rm UV}$ is the UV luminosity and $r$ is the distance between a stellar particle and a gas one. We estimate $L_{\rm UV}$ by integrating the SED of a stellar particle from $\lambda = 1000 - 5000~\rm \AA$, which is the range that radiation is efficiently absorbed by dust. \\

\noindent
{\it   $\bullet$ Hydrogen molecule dissociation}\\
In the case of First runs, we consider the dissociation process of hydrogen molecules due to Lyman-Werner (LW) feedback \citep{Johnson13}. Here we consider $\rm H_{2}$ dissociation and $\rm H^{-}$ detachment due to local radiation sources. The LW mean intensity is estimated by 
\begin{equation}
J_{\rm LW, 21} = \sum_{i=1}^{n} f_{\rm LW} \left( \frac{r_{\rm i}}{\rm 1~kpc}\right)^{-2} \left(  \frac{m_{\rm *, i}}{10^{3}~\Msun} \right),
\end{equation}
where $J_{\rm LW, 21}$ is described in unit of $10^{-21}~\rm erg \; s^{-1} \; cm^{-2} \; Hz^{-1} \; str^{-1}$,
$r_{\rm i}$ is the distance from $i$-th stellar particle to a target gas particle and $m_{\rm *, i}$ is the mass of $i$-th stellar particle. 
The normalization factor $f_{\rm LW}$ depends on the shape of SEDs. However, \citet{Sugimura17} showed that the feedback strength of young Pop II stars per unit mass was similar to that of Pop III stars \citep[see also][]{Agarwal16}. Therefore, unlike \citet{Johnson13}, we use the same value to both Pop III and II stars, and it is $f_{\rm LW}=15$.
The LW radiation can be attenuated locally due to self-shielding gas \citep[e.g.,][]{Draine96, Glover01, Wolcott-Green11, Wolcott-Green17, Luo20}.
 To take the self-shielding effect into account, we evaluate the column density over the local Jeans length as follows: 
\begin{equation}
N_{\rm H_{2}} = 2 \times 10^{15}~{\rm cm^{-2}}~ \left( \frac{f_{\rm H_{2}}}{10^{-6}}\right)
\left( \frac{n_{\rm H}}{10~\rm cm^{-3}} \right)^{1/2} \left( \frac{T}{10^{3}~\rm K} \right)^{1/2},
\end{equation}
where $f_{\rm H_{2}}$ is the fraction of $\rm H_{2}$, $n_{\rm H}$ is the hydrogen number density. 
Using the column density, we estimate the shielding factor based on \citet{Wolcott-Green11} as
\begin{equation}
\begin{split}
f_{\rm shield}(N_{\rm H_{2}}, T) = &\frac{0.965}{(1 + x/b_{5})^{1.1}}
+ \frac{0.035}{(1+ x)^{0.5}} \\
&~~~\times {\rm exp} \left[ -8.5 \times 10^{-4} (1+x)^{0.5} \right],
\end{split}
\end{equation}
where $x \equiv N_{\rm H_{2}}/5 \times 10^{14}~\rm cm^{-2}$ and $b_{5} \equiv b/10^{5}~\rm cm~s^{-1}$. 
Here $b$ is the Doppler broadening parameter, $b \equiv (k_{\rm B}T/m_{\rm H})^{1/2}$. 
Thus, we estimate the $\rm H_{2}$ dissociation rate ($\kappa_{\rm diss}$) by combining $J_{\rm LW, 21}$ and $f_{\rm shield}$ as $\kappa_{\rm diss} \propto f_{\rm shield} J_{\rm LW, 21}$. 
Once stars form in a halo, star formation in some nearby minihalos is suppressed due to the LW feedback \citep[e.g.,][]{Latif20}. 
As the halo mass increases or gas is metal-enriched, gas can collapse via metal cooling or hydrogen atomic cooling. 

\subsection{Supernova Feedback}

In this work, we consider supernovae (SNe) feedback via the injection of thermal energy into neighboring gas particles as described in \citet{DallaVecchia12}. Using random numbers, gas particles are chosen stochastically and heated up to $T=10^{7.5}~\rm K$.
The hot gas region pushes out the surrounding ISM due to the higher thermal pressure. 
This can lead to galactic-scale outflow if the thermal energy is converted to kinetic energy efficiently. 
The conversion rate depends on the local physical properties, e.g., gas density, clumpiness, metallicity \citep[e.g.,][]{Cioffi88, Kim15}.
\citet{DallaVecchia12} compared the sound crossing time with the cooling time, and derived the following maximum gas density for which the thermal energy is efficiently converted into the kinetic energy against radiative cooling losses:  
\begin{equation}
n_{\rm H} \sim 100~{\rm cm^{-3}} \left( \frac{T}{10^{7.5}~\rm K}\right)^{3/2} \left( \frac{m_{\rm g}}{10^{4}~\rm \Msun} \right)^{-1/2}.
\label{eq:nth}
\end{equation} 

Some star-forming regions can exceed the above critical density and suffer from the over-cooling problem. 
Also, in  regions with lower metallicity and lower gas density, the SN explosion energy is easier  
converted into  kinetic energy due to lower cooling rates \citep[e.g.,][]{Cioffi88, Thornton98}. 
Therefore, as introduced in S15, we consider a multiplication factor ($f_{\rm th}$) to the SN energy depending on local metallicity and gas density as
\begin{equation}
f_{\rm th} = f_{\rm th, min} + \frac{f_{\rm th,max} - f_{\rm th,min}}{1 + \left(\frac{Z}{0.1~\Zsun} \right)^{n_{\rm z}} \left( \frac{n_{\rm H, birth}}{n_{\rm H,0}}\right)^{-n_{\rm n}}},
\end{equation}
where $n_{\rm H, birth}$ is the gas density at which the star particle is formed, $n_{ Z}=n_{ n} = 2/{\rm ln}(10)$, and $n_{\rm H, 0} = 0.67~\rm cm^{-3}$, which were chosen after the comparison tests in S15. 
We here use the asymptotic values $f_{\rm th, max}=2.5$ and $f_{\rm th, min} = 0.3$. As discussed in S15, $f_{\rm th}$ can exceed unity. This is motivated by the additional feedback processes, not included in the simulations, e.g., stellar winds, cosmic rays, or if supernova yield more energy per unit mass than assumed here.  
Since we consider radiative feedback from young stars, we use a somewhat lower value of $f_{\rm th, max}$ than S15 ($f_{\rm th, max}=3.0$). 
\citet{Crain15} discuss that the dependencies  of the CSFRD and other properties of simulated galaxies on the choice of $f_{\rm th}$.  They concluded that the above model of $f_{\rm th}$ reproduced the observations of local galaxies well. 
The resolution of the PCR runs can allow a maximum density of $n_{\rm H} \sim 5 \times 10^{3}~\rm cm^{-3}$. Therefore, such a high-density region can still suffer from the over-cooling, although it is rare. 

\subsection{Black hole}

As galaxies evolve, massive black holes (BHs) are likely to form at the galactic centers \citep[e.g.,][]{Kormendy13}. 
Massive BHs can suppress star formation via radiative and kinetic feedbacks \citep[e.g.,][]{Dubois12}. 
Recent simulations show that star formation in massive galaxies can be suppressed by  BH feedback 
to reproduce the stellar-to-halo-mass ratio \citep[SHMR; e.g.,][]{Pillepich18}. 
To account for this we include BH feedback in our simulations.
%We put a BH with a mass of $10^{5}~\Msun/h$ at the galactic center if the halo once its mass exceeds $10^{10}~\Msun/h$. 
We replace the most high-density gas particle by a BH  with a mass of $10^{5}~\Msun/h$ in the halo once its mass exceeds $10^{10}~\Msun/h$. 
Gas accretion rate on the BHs is estimated based on the Bondi rate \citep{Bondi44}  using  100 neighbor gas particles as
\begin{equation}
\dot{m}_{\rm Bondi} = \frac{4 \pi cGM_{\rm BH}^{2} \rho}{(c_{\rm s}^{2} + v_{\rm rel}^{2})^{3/2}}
\end{equation}
where $v_{\rm rel}$ is the relative velocity between the BH particle and gas particle.  
As in S15, we consider a suppression factor due to angular momentum of gas, 
\begin{equation}
\dot{m}_{\rm acc} = \dot{m}_{\rm Bondi} \times {\rm min} \left( C_{\rm visc}^{-1}(c_{\rm s}/V_{\phi})^{3}, 1 \right)
\label{eq:dmbh}
\end{equation}
where $ C_{\rm visc}$ is a free parameter related to the viscosity of subgrid accretion disc \citep{Rosas-Guevara15}.
We set $C_{\rm visc} = 200 \pi$ which is same as in the AGNdT9 run in S15, which has been shown to reproducing the observed X-ray luminosity function well \citep{Rosas-Guevara16}.  BHs grow with the rate $\dmbh = (1-f_{\rm r}) m_{\rm acc}$ where $f_{\rm r} = 0.1$ is the radiative efficiency factor. 
The simulations suffer from resolving high-density gas within the Bondi radius. 
By taking the balance between the Bondi rate without the relative velocity and the Eddington accretion rate ($\dot{m}_{\rm Edd}  =  L_{\rm Edd} / (f_{\rm r}c^{2})$), we evaluate the gas density around a BH that would allow for Eddington accretion \citep{Park11, Yajima17b}
\begin{equation}
\begin{split}
    n_{\rm H} &\sim \frac{c_{\rm s}^{3}}{G \sigma_{\rm T} c f_{\rm r} M_{\rm BH}} \\
    &\sim 40~{\rm cm^{-3}} \left(\frac{M_{\rm BH}}{10^{5}~\Msun} \right)^{-1} \left( \frac{T_{\rm gas}}{10^{4}~\rm K}\right)^{1.5} \left( \frac{f_{\rm r}}{0.1}\right)^{-1}.
\end{split}
\end{equation}

Therefore some previous studies with low numerical resolutions had to introduce a boost factor to the accretion rate. 
On the other hand, the numerical resolutions of recent simulations can follow the accumulation of high-density gas around BHs, resulting in the efficient growth of BHs without the boost factor, e.g., as in S15.
Also, our simulations follow the growth of BHs without the boost factor and reproduce the formation of SMBHs in massive galaxies successfully, of which the masses distribute near the local relation between the BH and stellar-bulge mass.
In our fiducial model, we set the upper limit of the accretion rate as the Eddington limit, 
\begin{equation}
\dot{m}_{\rm Edd} = \frac{4 \pi G \Mbh m_{\rm p}}{f_{\rm r}\sigma_{\rm T}c}. 
\end{equation}

In the current resolution, it is difficult to follow the migration process of BHs due to  dynamical friction. 
We therefore  artificially model the migration of BHs toward the galactic centers by replacing them 
to the position of the potential minimum of neighbouring  particles. 
Once the BH settles at a galactic center, it starts to grow efficiently. 
Then, the growth can be self-regulated via feedback from the BH. 
In this work, we consider two types of feedback processes as described below. \\

\noindent{
$\bullet$ {\it Quasar mode feedback}} \\
The energy from an accretion disk is deposited into neighbouring gas particles thermally and gas particles are heated up to $T=10^{9} ~\rm K$. 
The released energy is estimated by $\Delta E = f_{\rm e} f_{\rm r} \dot{m}_{\rm acc} c^{2}$, 
where  $f_{\rm e}$ is the thermal coupling factor. 
Here we assume $f_{\rm e} = 0.15$ and $f_{\rm r} = 0.1$ for all simulations.    %for the large-scale simulations (PCR series). 
%In the cases of high-resolution simulations (BCG and First series), the conversion rate from thermal to kinetic energy can be high even in  high-density regions (see eq. (8)). Therefore, we use a somewhat lower value of $f_{\rm e} =0.1$ considering efficient thermal feedback. 
Unlike S15, we choose the nearest gas-particle and inject the thermal energy. 
Therefore, in case of continuous high gas accretion rates, the same gas particle can be selected as the target of the thermal feedback energy injection, likely resulting in heating up to $\gg 10^{9}~\rm K$. To avoid a gas particle getting to too high temperatures, alternative gas particles are selected in order of the distance from the BH, 
if thermal energy injection occurs continuously. 
 \\
 
 \noindent{
$\bullet$ {\it Radio mode feedback}}\\
As observed radio galaxies, supermassive black holes (SMBHs) with the mass $\sim 10^{9}~\Msun$ are likely to have impact on galactic scale via  jet-like kinetic feedback. 
We therefore, inject half of $\Delta E $ as kinetic energy and the other as thermal energy,
once the BH mass exceeds $10^{9}~\Msun$. 
We add the momentum to the gas kicked in 
the radio mode feedback to follow the direction of the angular momentum vector of neighbouring gas particles or the opposite direction, and the kick velocity is $3000~\rm km~s^{-1}$. 
The direction of the kick velocity is set along the angular momentum vector of surrounding gas ${\bold n_{1} } = {\bold L}/|L|$ or the inverse direction ${\bold n_{2} } = - {\bold L}/|L|$. 
The angular momentum is estimated from 100 neighbouring gas particles. 
We determine either direction via random numbers. 
Note that, we allow the hydrodynamical interaction of the kicked gas particles. Therefore, they can  thermalize via  shocks with the interstellar matter.
\\

\noindent{
$\bullet$ {\it Super-Eddington mode}}\\
Recent simulations show that disc winds can be launched due to the radiation from the inner parts of an accretion disc \citep[e.g.,][]{Murray95, Proga00, Proga04, Nomura20}. 
This disc wind can obscure the radiation from the accretion disc and generate an anisotropic radiation field. In the case of an  anisotropic radiation field, the gas accretion rate onto BHs can simply be proportional to the Bondi rate and not capped at the Eddington limit \citep[e.g.,][]{Netzer87, Wada12, Sugimura17}. 
Therefore, only for the run  BCG0spEdd, we allow  super-Eddington accretion, but set the maximum Eddington factor $f_{\rm Edd} = 5$. 
When the accretion rate exceeds the Eddington accretion rate, the radiative efficiency can be low due to the photon trapping in a slim disk \citep[e.g.,][]{Jaroszynski80}. %(e.g., Jaroszynski et al. 1980). 
We evaluate the luminosity of BHs based on a fitting formula \citep{Watarai00}, 
\begin{equation}
L = 
\begin{cases}
%&2 L_{\rm E} \left[ 1 + {\rm ln} \left( \frac{\dot{m}}{20}\right)\right] ~~~{\rm if}~~~ \dot{m} > 20\\
%&0.1 L_{\rm E} \dot{m} ~~~~~~~~~~~~~~~~{\rm if}~~~\dot{m} \le 20,
2.0 L_{\rm Edd} \left[ 1 + {\rm ln} \left( \frac{\dot{m}}{2.0}\right)\right] ~~~&{\rm if}~~~ \dot{m} > 2.0\\
 L_{\rm Edd} \dot{m} ~~~~~~~~~~~~~~~~&{\rm if}~~~\dot{m} \le 2.0,
\end{cases}
\end{equation}
where $\dot{m} \equiv \dot{m}_{\rm acc}/\dot{m}_{\rm Edd}$ is the gas accretion rate normalized by the Eddington accretion rate and $L_{\rm Edd}$ is the Eddington luminosity $L_{\rm Edd} = 4 \pi cG\Mbh m_{\rm p}/\sigma_{\rm T}$. 
%$\dot{m}_{\rm Edd} = \frac{4 \pi  G m_{\rm H} M_{\rm BH} }{ c \sigma_{\rm T}} $. 

\subsection{Post-processing radiative transfer}
To study the observational properties of simulated galaxies, we carry out post-processing radiative transfer calculations for specific snapshots. 
We use the multi-wavelength radiative transfer code {\sc $\rm ART^{2}$} \citep{Li08, Li20, Yajima12a}. 
This code is developed based on a Monte Carlo technique and calculates the transfer of photon packets through an adaptive refinement grid structure. The newest version of {\sc $\rm ART^{2}$} can handle continuum fluxes from stars and black holes, $\lya$ line from ionized hydrogen, atomic metal lines, and CO lines. Moreover, the code can make two-dimensional images of surface brightness for specific frequency ranges. 
Using the code, we reproduced successfully observational properties of high-redshift galaxies 
\citep{Yajima12c, Yajima13, Yajima14c, Yajima15a, Yajima15c, Arata19, Arata20}.  
We will model the observational properties of member galaxies of PCs in the next papers. 
In this work, we study the dust obscuring of massive galaxies in the PCs and infrared luminosities from the dust thermal emission. 
By considering the radiative equilibrium state, we estimate the dust temperature locally and the flux densities at far-infrared wavelengths.

The adaptive refinement grid structures for the radiative transfer simulations are set to resolve the minimum smoothing length ($0.1 \times$ gravitational softening). The physical properties of each grid are estimated from neighbouring SPH particles with a spline kernel function and a smoothing length. 
Our simulations initially set the number of the base grid as $N_{\rm base}=4^{3}$ and makes higher resolution grids if a cell around the grid contains more than $N_{\rm th}=16$ SPH particles.
Using the local metallicity, we model the dust density as $\rho_{\rm d} = 8 \times 10^{-3} \rho_{\rm gas} \left( Z/Z_{\rm \odot} \right)$. This relation is supported by observation of local galaxies \citep[e.g.,][]{Draine07}.
We cast $10^{6}$ photon packets, which satisfies our convergence tests (Appendix \ref{sec:appendix}) and generate good resolution SEDs. 

%----------------------------------------------------------------------
%
% Section 3:  Results
%
%----------------------------------------------------------------------

\section{Results}
\subsection{PCR runs}

%===============================================
\begin{figure*}
	\begin{center}
		\includegraphics[width=16cm]{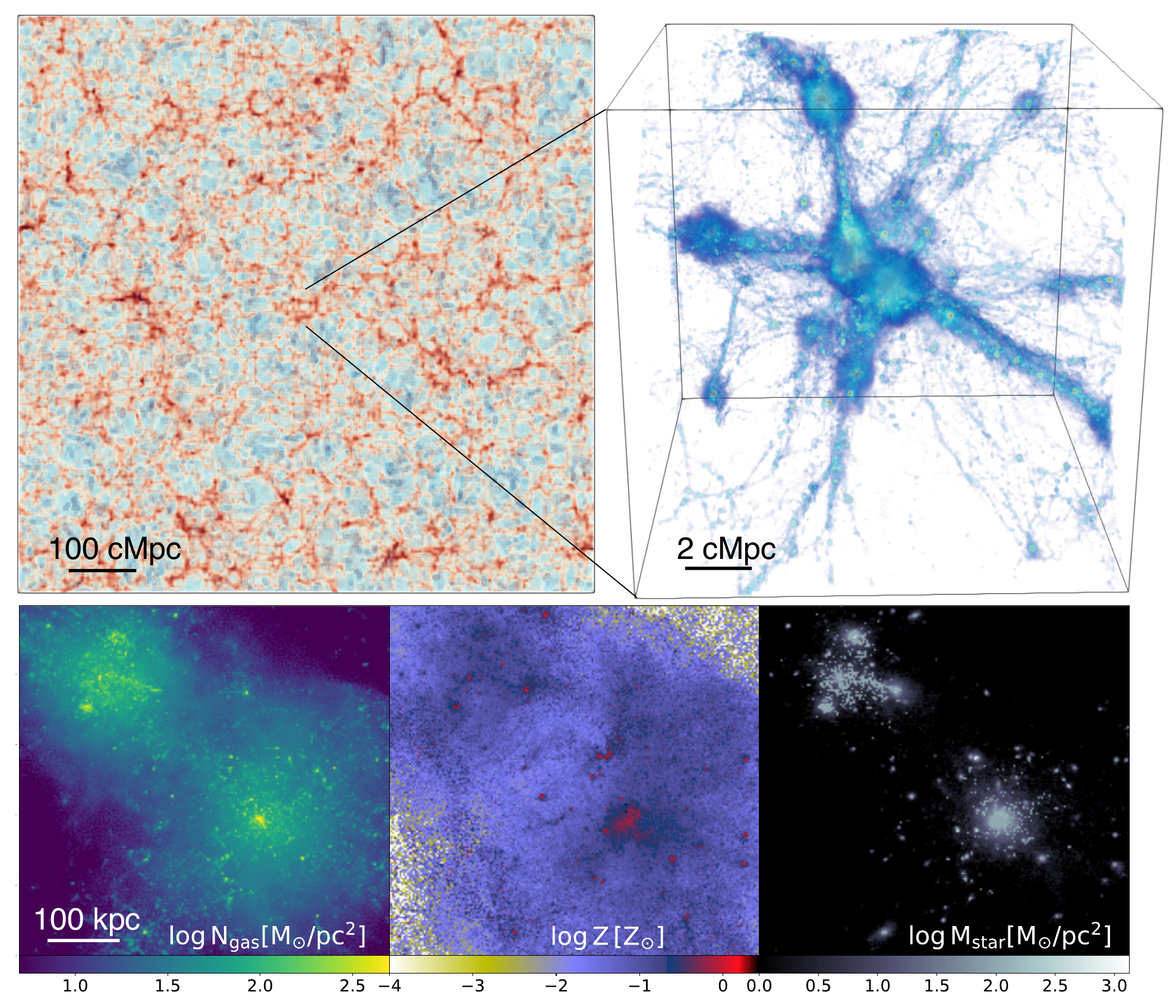}
	\end{center}
	\caption{	
	Upper left: The large scale structure of matter in the entire calculation box with $L=714 ~\rm cMpc$. 
	Upper right: Three-dimensional gas structure of the PCR0 region at $z=3$. 
	Lower panels: Gas column density (left), density-weighted metallicity (middle) and stellar surface density (right) of the most massive halo in PCR0.
		 }
	\label{fig:map}
\end{figure*}
%===============================================

Figure~\ref{fig:map} shows the distributions of gas, metallicity, and stars 
of the most massive halo in the PCR0 run and the large-scale structure at $z = 3$. 
As seen in the stellar distribution, the massive galaxies are undergoing a major merger. 
The total stellar mass and star formation rate in  
the halo are $2.5 \times 10^{12}~\Msun$ and $2679~\Msunyr$, respectively. 
The central parts of the galaxies already reach solar metallcity. 

\begin{figure*}
	\begin{center}
		\includegraphics[width=13cm]{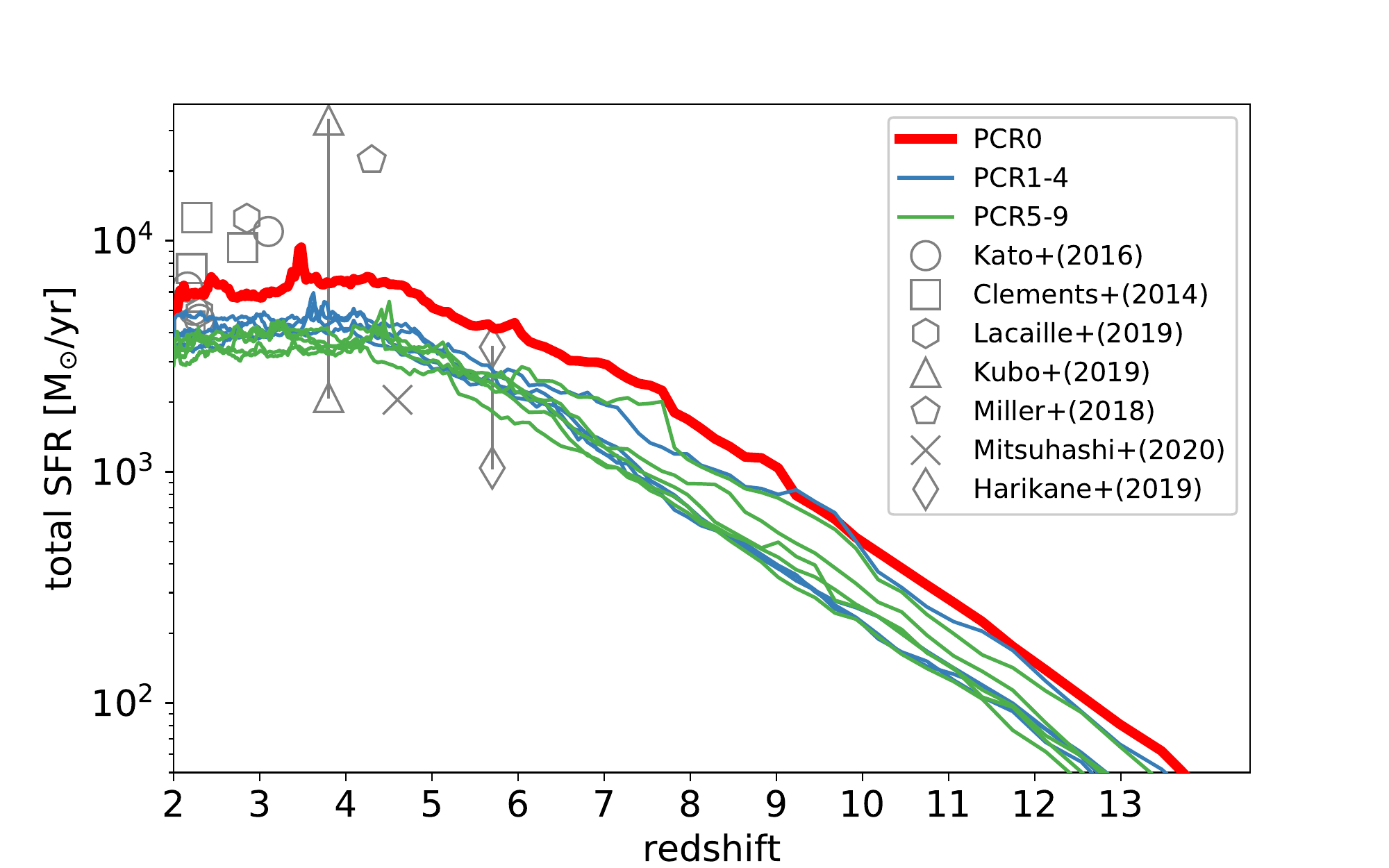}
	\end{center}
	\caption{	
	Total SFR within 10 cMpc from the centre of mass considering top 10 massive haloes in each zoom-in region. Red thick solid line shows the PCR0 run. Blue and green lines represent PCR1-4 and PCR 5-9 runs. Symbols are the observed total SFRs of protocluster candidates: 
	circle \citep{Kato16}, square \citep{Clements14}, hexagon \citep{Lacaille19}, triangle \citep{Kubo19}, pentagon \citep{Miller18}, cross \citep{Mitsuhashi20}, and dyamond \citep{Harikane19}. 
	The lower and upper values at $z=5.7$  assume that the fraction of associated submillimeter galaxies is 0.3 and 1.0, respectively, accounting for redshift uncertainty  
	\citep{Harikane19}. The upper and lower triangles repsent the values with and without AGN contribution \citep{Kubo19}. 
	%Gray diamonds are observed total SFRs of protocluster candidates: z=2.3 and 2.9 (Lacaille et al. 2019), z=4.3 (Miller et al. 2018) and z=4.6 (Mitsuhashi et al. 2020). 
	}
	\label{fig:totsfr}
\end{figure*}

Figure~\ref{fig:totsfr} presents the total SFRs within a radius of 10 cMpc, which corresponds to a typical Lagrange volume of PCs at $\gtrsim 2$ that makes clusters at $z \sim 0$ \citep[e.g.,][]{Chiang17}. 
Note that, in the case of massive galaxy clusters with $\gtrsim 10^{15}~\Msun$, the spatial distribution of the  building blocks at high-redshift is larger than 10 cMpc  \citep{Muldrew15, Lovell18}. Here, to simplify the analysis, we use 10 cMpc for all PCR runs. 
We estimate the total SFRs by using the SFR of each SPH particle at the time. 
The centre of the proto-cluster region  is chosen as the centre of mass of all massive galaxies with $\Mh \ge 10^{12}~\Msun$ in the zoom-in regions ($L=28.6~\rm Mpc$). The total SFRs monotonically increase with time at $z \gtrsim 4$ and then stall or   somewhat decrease from $z = 4$ to $2$. 
The evolution of the star formation histories differs from the cosmic star formation rate density (SFRD) as seen for the mean-field (MF) run or as derived from observations \citep{Madau14} in which the peak of SFRD is at $z \sim 1- 3$.
In the PC regions, massive galaxies form more frequently compared to the MF region. 
The massive galaxies consume gas via star formation earlier and the overall gas fraction becomes small at $z < 4$, resulting in a suppression of star formation activities. 
In addition, SMBHs form in the massive galaxies and hamper star formation via feedback. 
Note that, however, the star formation in massive galaxies is not quenched for a long time. Most of them can keep gas and maintain star formation, which will be discussed below.

Most of the PCR runs show a total SFR of $\sim 3000-5000~\Msunyr$ at $z = 2 - 4$. 
Only PCR0 exceed that with $6000~ \Msunyr$ at $z = 2- 4$ and achieves $9378~\Msunyr$ at $z=3.5$. 
The PCRs agree with the lower end of   
observed SFRs in PCs at $z \lesssim 3$ \citep{Lacaille19}, with some observed PCs 
being a factor of 2 - 3 higher. Note that, however, the estimate of the SFRs of observed PCs always suffers from the uncertainties of the dust temperature and the contribution of hidden AGNs. \citet{Kubo19} suggested that the total SFR of the PC they observe  at $z=3.8$ could be boosted due to the additional submillimeter flux from the dust-obscured AGN by $\sim 1~\rm dex$. %an order unity.
Also, in the observations, the field of view and the detection limits are not uniform. The dependency on the observed sky-area is discussed below. 

The PC regions show  total SFRs of $ > 1000 ~\Msunyr$ even at $z \sim 6 - 8$. 
Such high star formation rate will be accompanied with copious amounts of ionizing photons. 
Therefore these starburst regions are likely to induce cosmic reionization much earlier and make giant HII bubbles that have  high IGM transmission of $\lya$ lines from galaxies in the bubbles \citep{Yajima18}. 
We will investigate the relation between giant {\sc H ii} bubbles and the clustering of LAEs at the epoch of reionization in a follow-up study.

\begin{figure}
	\begin{center}
		\includegraphics[width=9cm]{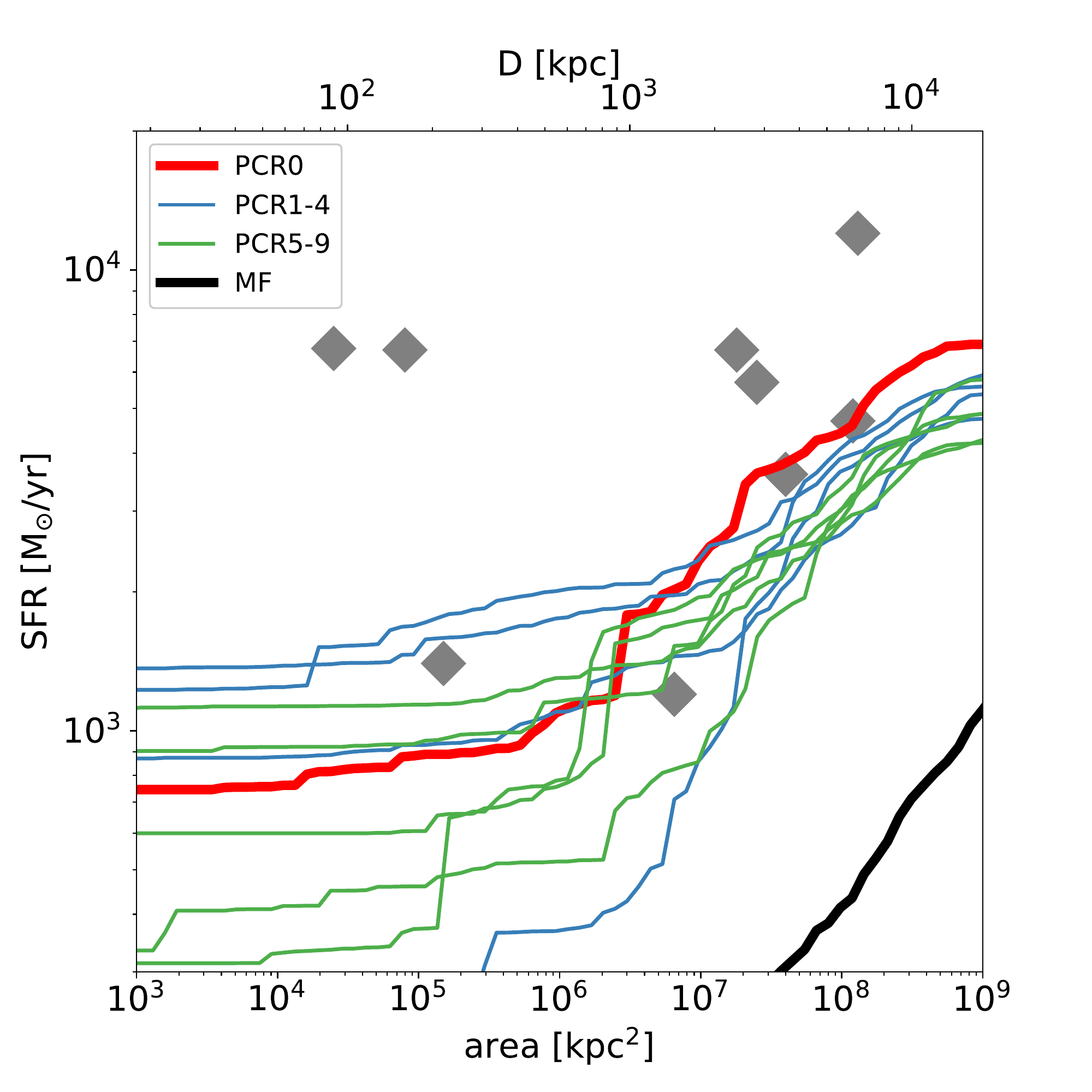}
	\end{center}
	\caption{	Cumulative SFR within a specific sky-area at $z=3$. The meaning of the colored solid lines is the same as in Figure~\ref{fig:totsfr}. The black solid line is the total SFR in the MF run.  Gray diamonds are the observed total SFRs of protocluster candidates shown in figure 2 in \citet{Miller18}. 	
		 }
	\label{fig:sfrdist}
\end{figure}

As shown in \citet{Miller18}, the concentration of starburst galaxies can be an important factor characterizing  PCs. Figure~\ref{fig:sfrdist} shows the cumulative SFR within a specific sky-area.
Here, we choose the most massive galaxy as the centre and integrate the SFR as a function of 2D radial distance
with the projection depth of $28.6~\rm cMpc$. 
The simulations show that the cumulative SFR  increases significantly at $10^{7} - 10^{8}~\rm kpc^{2}$. 
The most massive halo in PCR0 hosts five galaxies with ${\rm SFR} > 100~\Msunyr$ and there are seven starburst galaxies in the zoom-in region. Most other PCRs also have more than five galaxies with ${\rm SFR} > 100~\Msunyr$ in the zoom-in regions.
We find that the SFRs of all PCRs do not exceed $3000~\Msunyr$ at $\lesssim 10^{7}~\rm kpc^{2}$. 
This is because the typical separation distance between massive haloes is $\sim 1~\rm cMpc$ as seen in Figure~\ref{fig:pcrmap},  which requires at least a sky-area with $\sim 10^{7}~\rm kpc^{2}$ 
to include a  second massive halo with  high SFR. Some observed PCs also show a similar trend to the one reported in our simulations \citep[e.g.,][]{Casey15}. Whereas, even PCR0 cannot reach the high SFR of SSA22 which exceed $10^{4}~\Msunyr$ within $10^{8}~\rm kpc^{2}$. This may indicate that SSA22 is a more high-density rare peak or the SFR is overestimated because of hidden AGNs. 
We estimate total gas accretion rates onto BHs in the zoom-in regions that shows a diversity depending on the PC regions by a factor of $\sim 10$. Therefore, fluxes from some observed protoclusters might be boosted  due to AGNs.
Alternatively, the current simulation underestimates the SFR of massive haloes in PCs regions. Recently, \citet{Lim20} indicated that the SFR of simulated PCs increases significantly with the resolution of the simulations. 
However, note that, our simulation results do not change significantly at $z \lesssim 6$, according to test calculations with lower mass resolution (SPH particle mass is 8 times higher).
Furthermore, two protoclusters, SPT2349-56 \citep{Miller18} and S004224 \citep{Oteo17}, show highly concentrated star formation activity. 
These protoclusters reach $\sim 6000~\Msunyr$ even within $10^{5}~\rm kpc^{2}$ which is much higher than in other observed protoclusters and our simulations. For example, SPT2349-56 shows more than 10 starburst galaxies with $\rm SFR \gtrsim 100~\Msunyr$ coexisting within a small area. 

\begin{figure}
	\begin{center}
		\includegraphics[width=9.5cm]{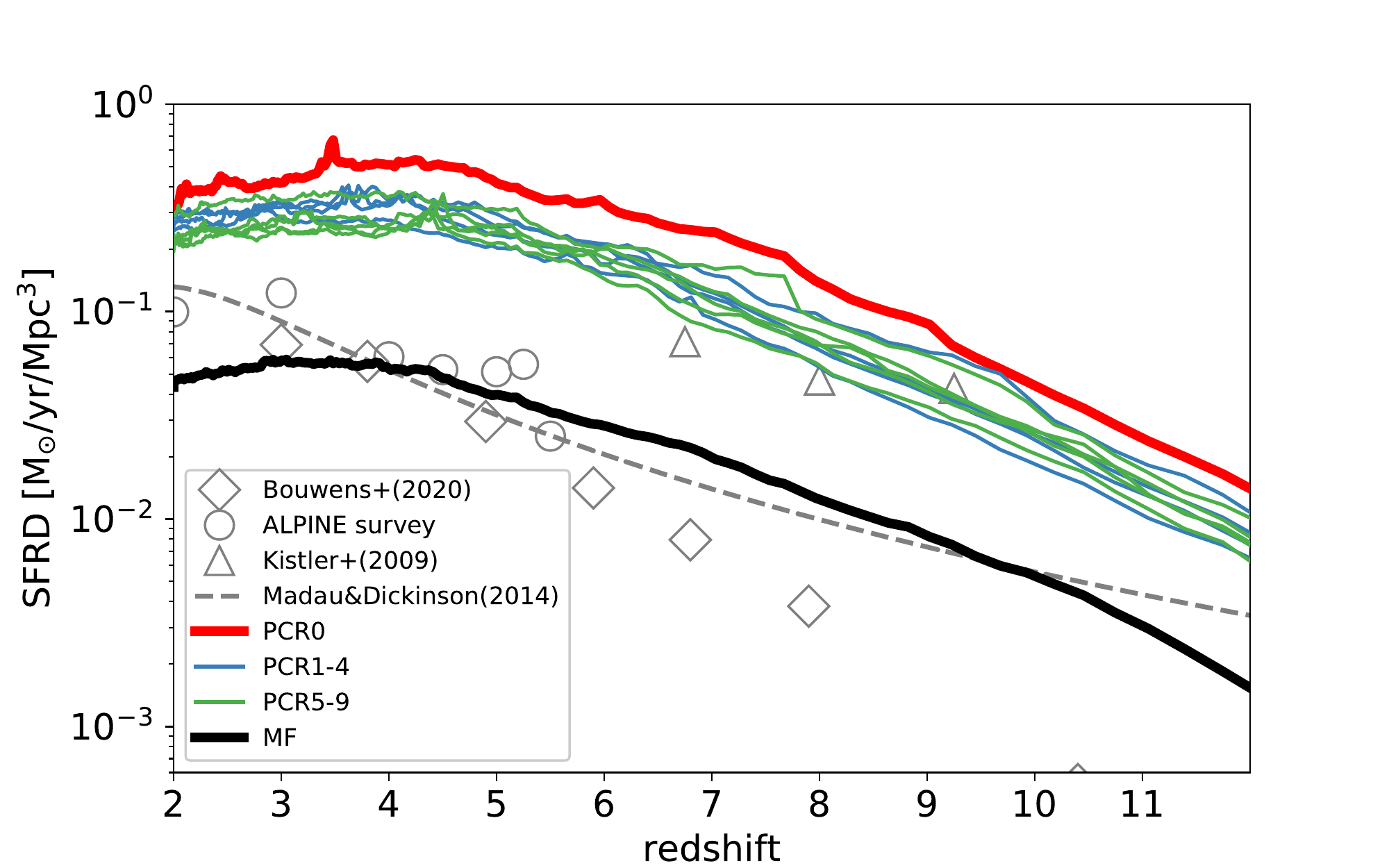}
	\end{center}
	\caption{		
		Cosmic star formation rate density. 
		The meaning of the different lines is the same as in Figure~\ref{fig:sfrdist}. 
		Open symbols show the observational data:
		diamonds from \citet{Bouwens20}, 
		circles ALMA ALPINE survey \citep{Khusanova20, Gruppioni20, Loiacono20}, 
		%crosses from Rowan-Robinson et al. 2016,
		triangles  from \citet{Kistler09}.
		The black dashed line shows the extrapolated fitting function for UV-selected galaxies at $z\leq 10$ derived in \citet{Madau14}. 
				 }
	\label{fig:sfrd}
\end{figure}

Cosmic star formation rate densities (SFRD) are presented in Figure~\ref{fig:sfrd}. 
As stated in Section \ref{sec:f22}, we use galaxies only in the inner regions with a volume of $(25.7~\rm cMpc)^{3}$ for estimates of statistical properties as SFRD, stellar mass function, main sequence and so on.
The SFRD of the MF run roughly matches the observations.  Earlier work has shown that the SFRD is regulated by  SNe feedback \citep[e.g.,][]{Schaye10}. As structure formation proceeds, haloes grow via mergers and matter accretion. Therefore the total star formation rate in the simulation volume increase. 
As the redshift decreases, the halo growth rate decreases gradually, and gas in galaxies is consumed by star formation, resulting in the plateau of SFRD $z \sim 2 - 3$. 
At $z \gtrsim 6$, there is large uncertainty in observed SFRDs. 
\citet{Oesch15} indicates that the SFRD drops down significantly at $z > 6$, 
while a recent survey of dusty star-forming galaxies with ALMA shows a higher SFRD \citep{Gruppioni20, Khusanova20}. The SFRD of the MF run lies between reported results from galaxy observations in the UV and rest frame infrared. 
The peak of SFRD of the MF run is somewhat earlier than that derived in \citet{Madau14}. 
Note that, the SFRD in simulations sensitively depends on the feedback model and resolution as shown in \citet{Schaye10} , because low-mass haloes are significant contributors. 
In the MF run, the impact of AGN feedback is secondary, we confirm that it reduces SFRD at $z \lesssim 3$ by at most a factor of 2 from a test calculation without  AGN feedback. 
Therefore,  SN feedback can play a role in shaping the SFRD. If future observations will determine the SFRD at $z > 4$ more precisely, it will constrain  SN feedback models in simulations tightly.

SFRDs in the  PCR runs are higher than that of the MF run by a factor of $\sim 3 - 5$. 
These differences are higher than the differences of total matter mass included in haloes in the zoom-in regions. 
In the overdense regions, more massive haloes form, and the halo number density is larger than in the mean-density field, leading to higher SFRDs. The shapes of the SFRDs of the PCR 1-4 runs are similar to that of the MF run, with the only difference that the normalisation is higher. On the other hand, PCR0 shows a slight decreases from $\sim 0.4~ \rm \Msun~yr^{-1}~Mpc^{-3}$ at $z = 4$ to $\sim 0.3~ \rm \Msun~yr^{-1}~Mpc^{-3}$ at $z=2$. 
This is due to AGN feedback. Some massive haloes host SMBHs with $\sim 10^{9}~\Msun$ at $z \lesssim 5$ which suppress star formation. The total BH mass in PCR0 is $7.3 \times 10^{10}~\Msun$ and higher than other PCR runs by a factor of $\sim 2-5$.

%===============================================

%===============================================

\begin{figure}
	\begin{center}
		\includegraphics[width=9cm]{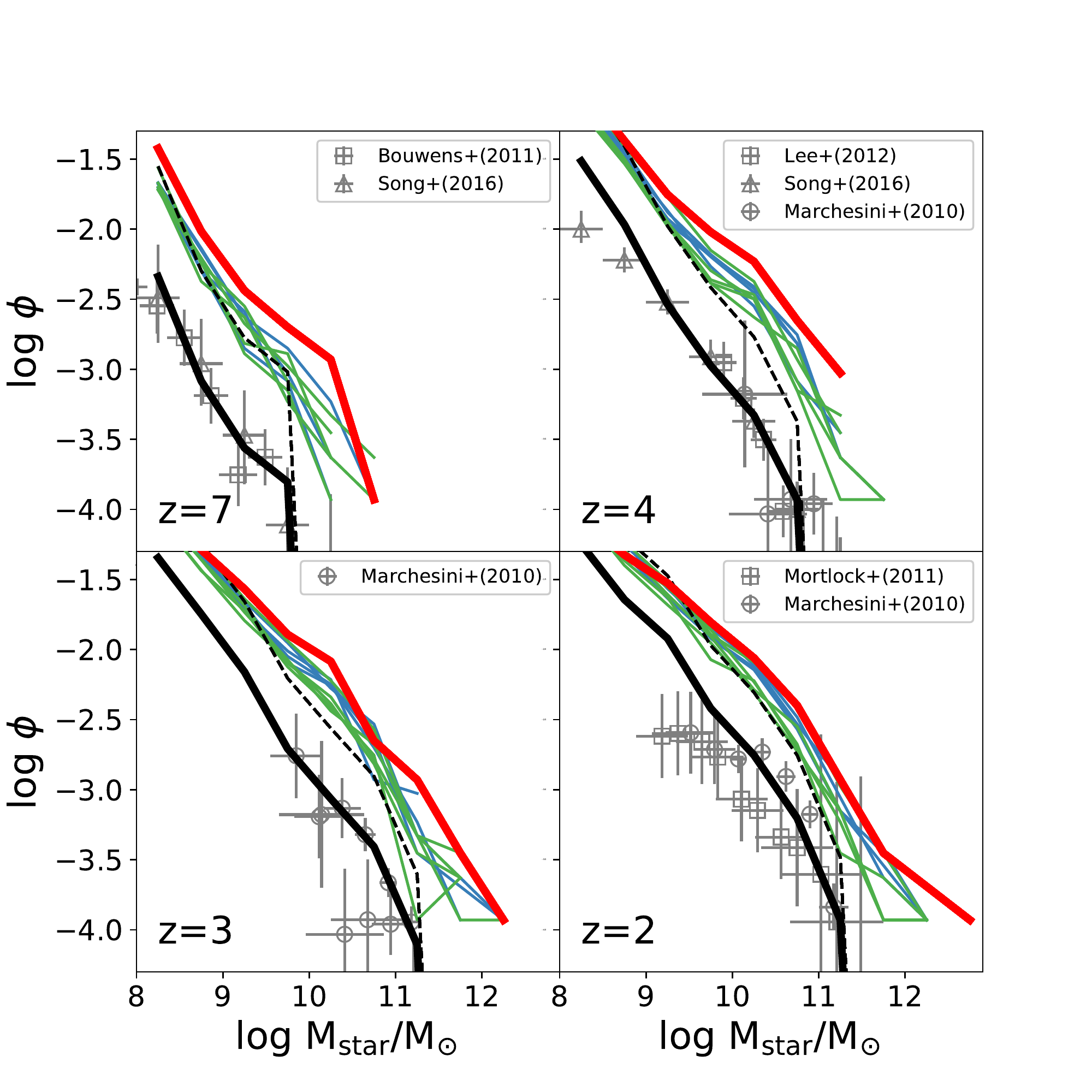}
	\end{center}
	\caption{	Stellar mass functions of MF and PCR runs at $z=7, 4, 3$ and $2$. Line types are the same as in  Figure~\ref{fig:sfrdist}.
    Black dashed line are the mass functions of MF scaled by mass ratios of total matter enclosed in haloes with $\Mh \ge 10^{10}~\Msun$ between MF and PCR0: 2.8 ($z=2$), 3.2 ($z=3$), 3.6 ($z=4$) and 6.1 ($z=7$). 
	Open symbols show the observed stellar mass functions: z=7 \citep[open squares:][]{Bouwens11b}, \citep[open triangles:][]{Song16a}; z=4 \citep[open circles:][]{Marchesini10}, \citep[open squares:][]{Lee12}, \citep[open triangles:][]{Song16a}; z=3 \citep[open squares:][]{Marchesini10}; z=2 \citep[open squares:][]{Mortlock11b}, \citep[open circles:][]{Marchesini10}.  
		 }
	\label{fig:stmf}
\end{figure}

% stellar mass function
Figure~\ref{fig:stmf} shows stellar mass functions at $z=2, 3, 4$ and $7$. We consider the total stellar masses of all galaxies identified by {\sc subfind}.
The MF run successfully reproduces the observed stellar mass functions at $z=2-7$. 
This indicates that the sub-grid models in our simulations are tuned reasonably \citep[see e.g. also][]{Cullen17}. 
\citet{Furlong15} also showed the  stellar mass functions from the EAGLE simulations that matched the observations at $z < 6$.  At $z \gtrsim 6$, their results for $\Mstar \gtrsim 10^{9}~\Msun$ are somewhat lower than our results or the observations. However, note that, there is still uncertainty in the stellar mass function at high redshifts. In addition, recently \citet{Wang19} suggested that a part of star-forming galaxies could be missed by UV-optical selection. Future multi-wavelength observations  will provide the stellar mass function and SFRD at high-redshifts more accurately and constrain theoretical models.  

The PCR runs always show stellar mass functions with large normalization $\phi$ that is higher by a factor $\sim 2-10$.  
%The total matter masses included in all haloes in PC runs are higher than MF by a factor of $\sim 2-3$. As a reference, we add the stellar mass functions of MF boosted by a factor of 2.5 artificially. 
The total matter masses included in all haloes with $\Mh \ge 10^{10}~\Msun$ in PCR0 are higher than MF by a factor of 2.8 ($z=2$), 3.2 ($z=3$), 3.6 ($z=4$) and 6.1 ($z=7$). As a reference, we add the stellar mass functions of MF boosted by the ratios of the total matter masses artificially. 
At $\Mstar \lesssim 10^{9}~\Msun$, it is similar to  the PCR runs. On the other hand, the number densities of massive galaxies of the PCR runs are larger apparently. 
This indicates that massive haloes form earlier at the cores of PCs and the star formation proceed rapidly.
Recently, \citet{Ando20} investigated the stellar mass function of the observed PC at $z \sim 2$ and it shows the excess at ${\rm log}(\Mstar/\Msun) > 10.5$ compared to the mean-density field. Our results match the observation. 

%Main sequence
The relation between SFR and stellar mass is used as a ruler of star formation activity. 
As in \citet{Pillepich18, Pillepich18b}, we estimate the gas/stellar mass and SFR within $2 \times r_{0.5}$,
where $r_{0.5}$ is the half mass radius of stars in the most massive member galaxy in a halo.
These physical quantities will also be used in next figures.
Figure~\ref{fig:mainseq} shows the SFRs as a function of stellar mass. 
We find most galaxies distribute along the observed main-sequence lines even in the PC regions at $z \gtrsim 2$. 
This trend was also reported by the  FLARES project for $z \ge 5$ \citep{Lovell21} and DIANOGA simulations \citep{Bassini20}.
\citet{Hayashi16} suggested that observed massive galaxies in a protocluster at $z=2.5$ were on the main-sequence \citep[but see,][]{Shimakawa18}.
Also, this is in agreement with results presented by \citet{Sparre15} who showed that most galaxies distributed along the main sequence at $z > 1$ in their simulation and that massive galaxies with $\Mstar \gtrsim 10^{11}~\Msun$ only get quenched at $z < 1$.
Note that the total stellar masses of some galaxies identified by {\sc subfind} are somewhat higher than the values estimated by the above method. However, the trend in the figure does not change significantly. The values can move to higher SFR and stellar mass along the main-sequence line in the case of using the stellar mass of the galaxies slightly.
On the other hand, observations indicate some massive galaxies should be quenched even at $z \sim 2$ \citep{Daddi05, Tacchella15, Tanaka19, Esdaile20}. 
%The quenching of star formation in massive galaxies can be induced by AGN feedback. In the current simulations of galaxy formation, the gas structure near the central BHs and their feedback are not well resolved. Therefore, various sub-grid models for gas accretion and feedback have been developed \citep{Dubois12, Schaye15, Nelson18}, and those models for massive galaxies are still under debate. 
However,  high-redshift passive galaxies are still quite rare. Therefore, the limited volume of our simulations may not be enough to reproduce such passive galaxies.

We find that the distribution of SFR in the PCR runs does not differ from that in the MF run. 
This suggests that star formation activity may not be sensitive to the environment and instead regulated locally. 
At $\Mstar \gtrsim 10^{10}~\Msun$, the dispersion in SFRs becomes large. 
Part of the massive galaxy population starts to deviate to lower SFRs with respect to the main sequence by more than $ 1$ dex. In our model, BHs rapidly grow in hosts with  $\Mstar \gtrsim 10^{10}~\Msun$ (see figure~\ref{fig:bhmass}). Therefore  BH feedback can evacuate gas from galaxies and suppress star formation.  
Looking at the feedback energy,  gas accretion at the Eddington limit onto a BH of $\Mbh = 10^{8}~\Msun$, 
generates $5.0 \times 10^{11}~L_{\rm \odot}$ in our model ($f_{\rm r} f_{\rm e} = 0.015$). 
That is much higher than the energy injection rate $\sim 0.8 \times 10^{10}~\L_{\rm \odot}$ from SNe for a ${\rm SFR} \sim 100~\Msunyr$ which is typical  for galaxies with $\Mstar \sim 10^{11}~\Msun$. 
Then, as the haloes grow, they can hold gas against feedback and form stars, resulting in SFRs near the main-sequence line.
We note that the quantitative results are likely to depend on the energy deposition rate from BHs. More efficient BH feedback can suppress star formation in massive galaxies \citep{Nelson18}. Recent observations indicated passive galaxies already formed even in the early Universe \citep{Glazebrook17, Mawatari20}, although they are rare. To understand the diversity of massive galaxies, BH/SN feedback should be investigated further in future studies. 

If the angular resolution of observations is not high, the entire region of a halo can be observed. 
Therefore we also evaluate the total SFR and stellar mass of haloes in the PCR0 run. 
The most massive halo in the PCR0 shows a total SFR of $2679~\Msunyr$ and $\Mstar = 2.5 \times 10^{12}~\Msun$, which corresponds to bright SMGs at $z \sim 3$ as seen in figure~\ref{fig:fesc}.
Recent ALMA observations have revealed multiple components in bright SMGs detected with SCUBA and suggest that multiple dusty star-forming galaxies are hosted in  massive haloes \citep{Simpson15}. 
Our simulations suggest that multiple SMGs resolved by ALMA can be hosted in a common massive halo that is a very bright SMG identified by a single-dish submillimeter telescope, e.g., SCUBA-2,  ASTE \citep[e.g.,][]{Tamura09}.

In our simulations, even active star-forming galaxies are distributed within $\sim 0.5$ dex from the main-sequence line. On the other hand, some observed SMGs showed $\sim 1$ dex higher SFRs at a specific stellar mass ($\sim 10^{11}~\Msun$) than the main-sequence. Because of the limited numerical resolutions, we force the polytropic equation of state to ISM once the local density exceeds the threshold  for star formation ($n_{\rm H} \sim 0.1~\rm cm^{-3}$) to avoid the artificial fragmentation. 
While this model can keep a stable galactic disc and reproduce the observed galaxy sizes \citep{Furlong15}, the violent disc instability may not be followed. Therefore, if a high sSFR is induced by a disc instability, we need to relax  forcing particles onto the EOS via increasing the numerical resolution. 

%stellar mass - halo mass
Figure~\ref{fig:smhm} shows the stellar-to-halo mass ratios (SHMRs). 
The SHMRs increase monotonically at $\Mh \lesssim 10^{12.5}~\Msun$
and then decrease toward the massive end. 
The star formation in low-mass haloes is suppressed due to the SN feedback. 
Therefore, SHMRs of low-mass halos with $\Mh \sim 10^{11}~\Msun$ can change with the parameter $f_{\rm th}$ by a factor of few (see also, Crain et al. 2015).
Given that a weaker SN feedback model or $f_{\rm th}=1$ is used, the SFR and stellar mass of low-mass haloes increase significantly.
As the halo mass increases,  haloes can hold the gas against SN feedback and allow efficient star formation, resulting in the high SHMRs $\gtrsim 10^{-2}$ at $\Mh \sim 10^{12}~\Msun$. 
In massive haloes with $\Mh \gtrsim 10^{13}~\Msun$,  the gas fraction of galaxies decreases, and SMBHs can provide additional strong feedback. Therefore, the SHMRs of massive galaxies in the PC regions become smaller $\rm SHMR \lesssim 10^{-2}$.

%Gas fraction
The ratio of gas mass to total baryon mass (gas+stars) is presented in figure~\ref{fig:fgas}. 
The gas mass fraction ($\fgas$) monotonically decreases as the stellar mass increases. 
We find $\fgas \gtrsim 0.8$ at $\Mstar \sim 10^{8}~\Msun$ and $\fgas \lesssim 0.4$ at $\Mstar \gtrsim 10^{11}~\Msun$. This implies that the gas in galaxies is consumed by star formation at a higher rate than the gas fueling.
Also, in massive haloes,  AGN feedback can contribute to expel the gas, and the cooling time of halo gas is long, suppressing the recovery of gas. 
\citet{Troncoso14} estimated the gas content of galaxies at $3 \le z \le 5$, including the SSA22 region, 
by combining SFRs within specific radii and the Schmidt-Kennicutt relation. 
We estimate the gas fraction using the gas and stellar mass within $2 \times r_{0.5}$.
Our results match the observations.
Note that, however, the observations consider cold neutral gas alone. Our simulations cannot distinguish cold gas alone and include hot ionized gas due to resolution limitations, and the pressure floor using the polytropic equation of state with $\gamma = 4/3$ is used. Therefore our estimation of $f_{\rm gas}$ can be somewhat higher than if considering  cold gas alone.
We also estimate $\fgas$ by using the total stellar and gas masses in haloes, i.e., within a virial radius. 
It shows the high values of $\gtrsim 0.8$, irrespective of the stellar mass as seen by the open circles and triangles. 
These discrepancies of $f_{\rm gas}$ between haloes and galaxies (star-forming regions) imply that most of the gas keeps being trapped in massive haloes even if they are pushed by the feedback. 
The cooling time of the halo gas can be estimated as
\begin{equation}
\begin{split}
&t_{\rm cool} = \frac{3 k T}{2 n_{\rm H}^{2} \Lambda (T)} \\
 &= 3.3 ~{\rm Gyr} ~\left( \frac{T}{10^{6}~\rm K}\right)
 \left(\frac{n}{10^{-3}~\rm cm^{-3}} \right)
 \left(  \frac{\Lambda (T)}{10^{-23}~\rm erg~s^{-1}~cm^{3}}\right)^{-1}
\end{split}
\end{equation}
where $\Lambda (T)$ is the radiative cooling rate. 
If the temperature of the halo gas is close to the virial temperature, the cooling time of massive haloes with $\Mh \gtrsim 10^{12}~\Msun$ is longer than the depletion time while on the main-sequence:
\begin{equation}
\begin{split}
t_{\rm dep} &\sim \frac{M_{\rm disc}}{\rm SFR} \\
&\sim 1.0~{\rm Gyr}~\left( \frac{f_{\rm disc}}{0.05}\right)
\left( \frac{\Mh}{10^{12}~\Msun}\right) \left( \frac{\rm SFR}{250~\Msunyr}\right)^{-1},
\label{eq:tdep}
\end{split}
\end{equation}
where $f_{\rm disc}$ is the mass ratio of gaseous disc to the halo mass.
Once galaxy merger or disc instability occurs, the disc quickly looses angular momentum, resulting in gas flow to the galactic center. 
In that case,  SFRs are likely to be proportional to $C_{*}M_{\rm disc}/t_{\rm dyn}$,  where $C_{*}$ is the conversion efficiency from the inflow rate to SFR and
$t_{\rm dyn}$ is the dynamical time of the galactic disk which can be evaluated as
\begin{equation}
\begin{split}
t_{\rm dyn} &\sim \frac{\lambda R_{\rm vir}}{V_{\rm \phi}} \\
&\sim 1.5 \times 10^{-2}~{\rm Gyr}~\left( \frac{\lambda}{0.05}\right)
\left( \frac{\Mh}{10^{12}~\Msun} \right) \\
&~~~~~~~~~~~\times \left(\frac{1+z}{4} \right)^{-1}
\left( \frac{V_{\rm \phi}}{250~\rm km~s^{-1}}\right)^{-1},
\end{split}    
\end{equation}
where $\lambda$ is the halo spin parameter and $V_{\rm \phi}$ is the rotation velocity of the galactic disc. The consumption time scale of the gas is estimated as $\sim t_{\rm dyn}/C_{*}$, and it becomes shorter than $t_{\rm dep}$ if $C_{*} > 1.5\times10^{-2}$.
Thus, in the case of massive haloes, the cooling time scale 
can be longer than the time scale for consumption by star formation. 
Therefore, once the gas in the galactic disc is expelled into the halo via  stellar or AGN feedback,
the halo gas is likely to be hampered to accrete onto the star-forming regions due to the thermal pressure support if the radiative cooling is inefficient.  
This can induce the large discrepancy of $\fgas$ seen between galaxies and haloes at $\Mstar \gtrsim 10^{10}~\Msun$.  

On the other hand, some massive galaxies show  high gas fraction with $\fgas \gtrsim 0.6$. 
As shown in Figure~\ref{fig:pcrmap}, massive haloes in the PCs form at the crossing of  large-scale filaments.
Therefore, the IGM filaments can feed  massive galaxies with gas efficiently, leading to the formation of  gas-rich massive galaxies.
We will investigate the detailed motion of inflow and outflow of gas from  massive haloes in future work.

%Metallicity
Figure~\ref{fig:metallicity} presents gas phase metallicities. 
We measure the metallicity by using gas particles within $2 \times r_{0.5}$. 
As star formation proceeds, metals ejected from SNe are accumulated in galaxies. 
Therefore, the metallicity increases with the stellar mass monotonically. 
In low-mass haloes, a part of the metal-enriched gas can be expelled due to the galactic winds which results in a steep mass dependency of the metallicity. 
The metallicity reaches $\sim 0.5 \times$ solar abundance at $\Mstar \sim 10^{10}~\Msun$. 
At $\Mstar > 10^{10}~\Msun$, the metallicity becomes almost constant within $Z \sim 0.5 - 1 ~\Zsun$. 
This trend is similar to reported observed relations \citep{Maiolino08, Mannucci09, Onodera16}. 
Note that, however, some massive galaxies show somewhat lower metallicities.
These galaxies are gas-rich as shown in Figure~\ref{fig:fgas}, which indicates that they are fueled by low-metallicity gas, likely from IGM filaments.

The metallicities of galaxies with $\Mstar \lesssim 10^{9}~\Msun$ are somewhat higher than the observations. This is likely due to the arbitrary regions of measuring the metallicity in the simulations. For example, \citet{Shimizu14} took into  account the metallicities  weighted by the  local ionizing photon emissivities. 
If we consider wider regions, the metallicity at specific stellar-mass decreases 
because the gas metallicity becomes lower as the distance from the galactic center increases. 
As a reference, we also estimate the metallicity by using all gas particles in a halo. 
In that case, the metallicity becomes lower than the case using $2 \times r_{0.5}$ by a factor of 2-5, 
and the difference increases with the stellar mass.  
Besides, the metal distribution sensitively depends on the feedback model. We will study the relation between the metal distribution and the feedback models in future work. 

In addition, this might suggest that the observed SEDs with metal lines reflect gas at $> 2\times r_{0.5}$. 
Future missions with PFS on the Subaru telescope will investigate the radial distribution of metals using metal absorption lines in SEDs of background galaxies. 
The comparison of our simulations with future observation will allow understanding the origin of the discrepancies reported above.

%===============================================

\begin{figure}
	\begin{center}
		\includegraphics[width=9cm]{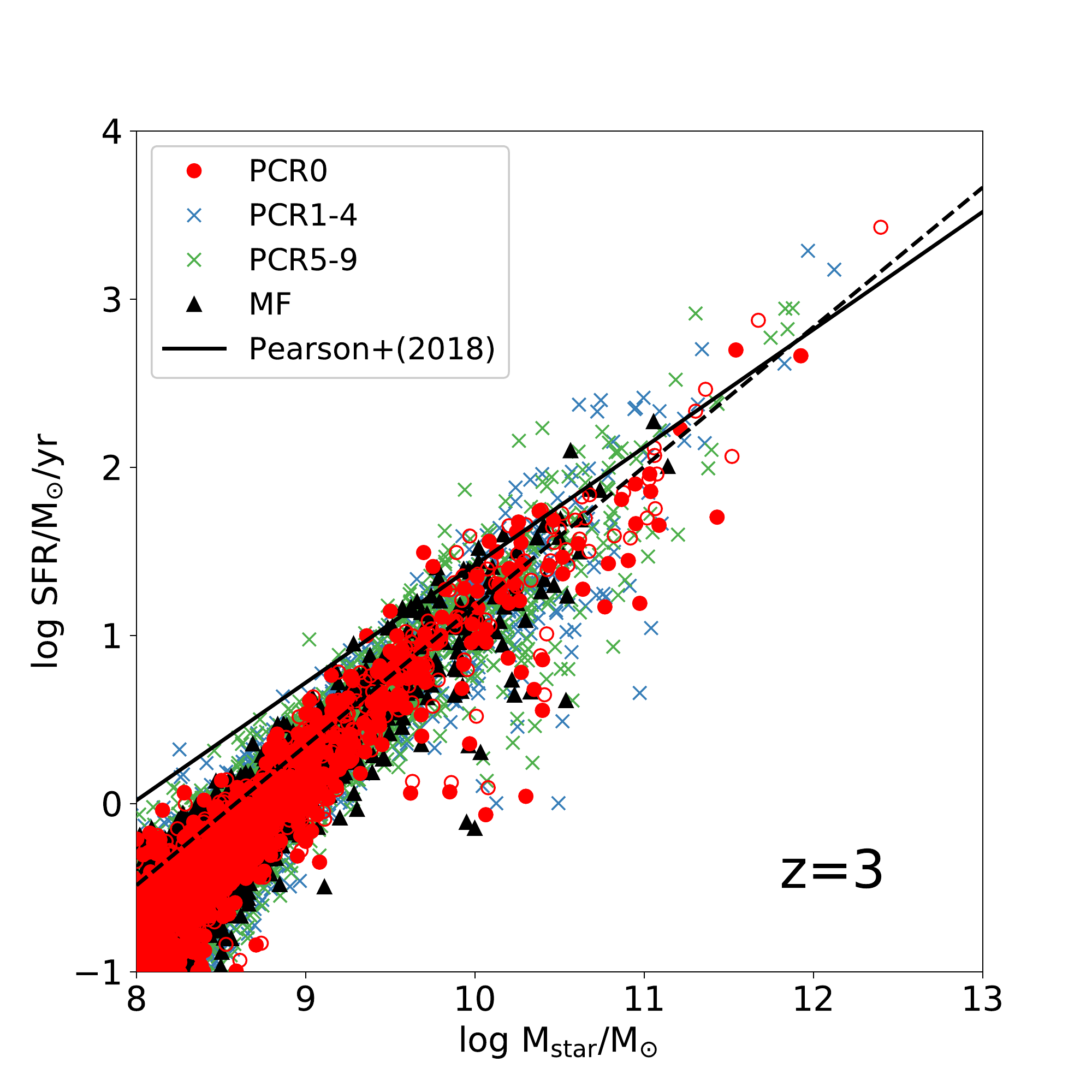}
	\end{center}
	\caption{	
	Star formation rates of galaxies as a function of stellar mass. 
	Different symbols represent each run: PCR0 (filled red circles), 	
	PCR1-4 (blue crosses), PCR5-9 (green crosses), and MF (filled black triangles). 
	The stellar mass and SFR are estimated within $2 \times r_{\rm 0.5}$ where
	$r_{0.5}$ is the half stellar mass radius of a most massive galaxy in a halo. 
	Open red circles show the case using total stellar mass and SFR in a halo. 
	Black dashed and solid lines show the relations of observed galaxies at $z = 2.3 - 2.9$ and
	$z = 2.9 - 3.8$ derived in \citet{Pearson18}. 
		 }
	\label{fig:mainseq}
\end{figure}

\begin{figure}
	\begin{center}
		\includegraphics[width=9cm]{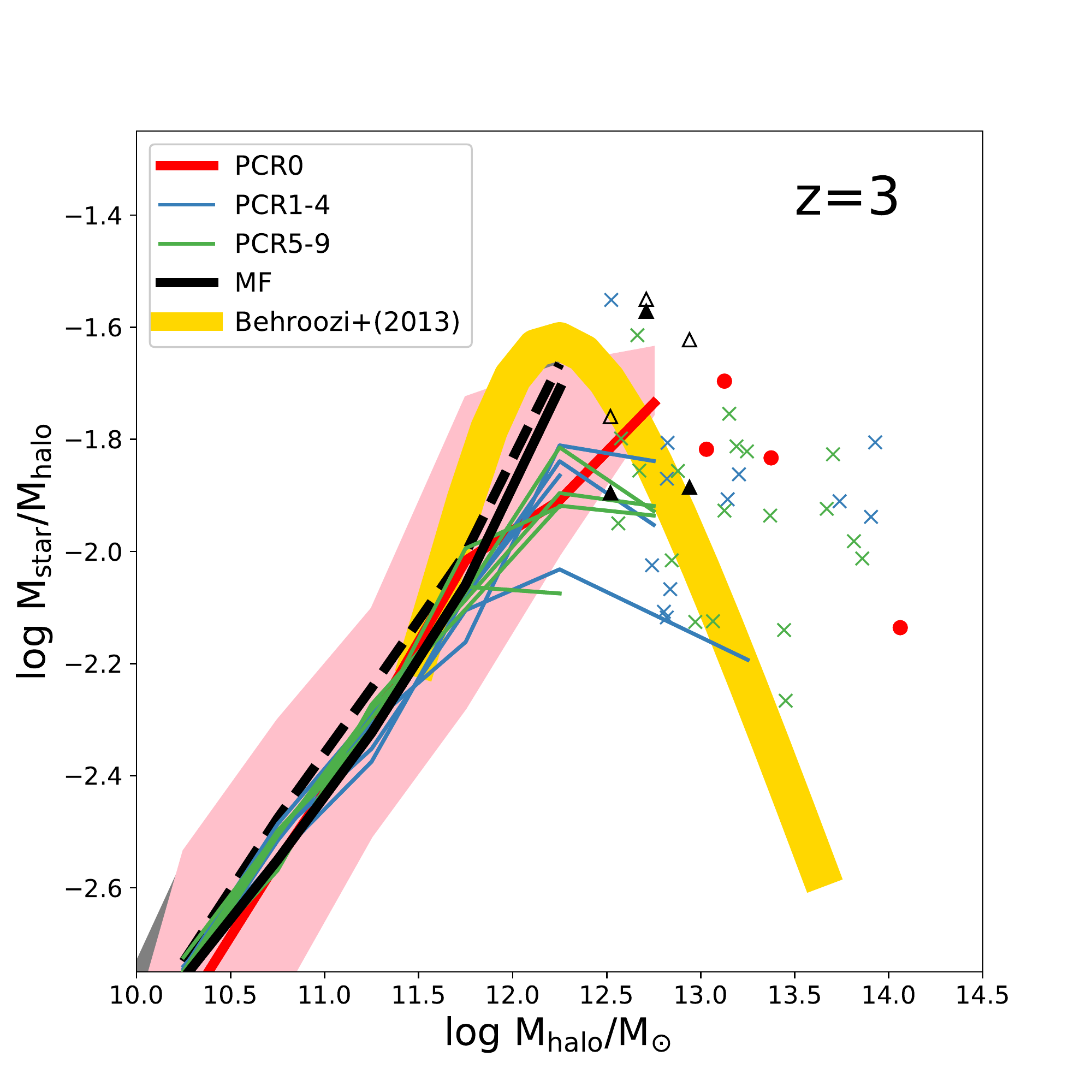}
	\end{center}
	\caption{	
	Stellar to halo mass ratio as a function of halo mass. The bin size is $\Delta {\rm log} \Mh/\Msun = 0.25$. 
	Lines represent the median values in each bin.  Line types are the same as in Figure~\ref{fig:sfrdist}.
	If the number of galaxies in a bin is smaller than five, the values of the galaxies are shown as symbols. 
	Different symbols show different runs, same as in figure \ref{fig:mainseq}. 
	The stellar mass is estimated within $2 \times r_{\rm 0.5}$. 
	The pink and gray shades show the quartiles (25 - 75 percent) in each bin in PCR0 and MF runs. 
	Black dashed line and open triangles are based on the total stellar mass in haloes in MF run. 
	The yellow thick curve is taken from \citet{Behroozi13}. 
		 }
	\label{fig:smhm}
\end{figure}

\begin{figure}
	\begin{center}
		\includegraphics[width=9cm]{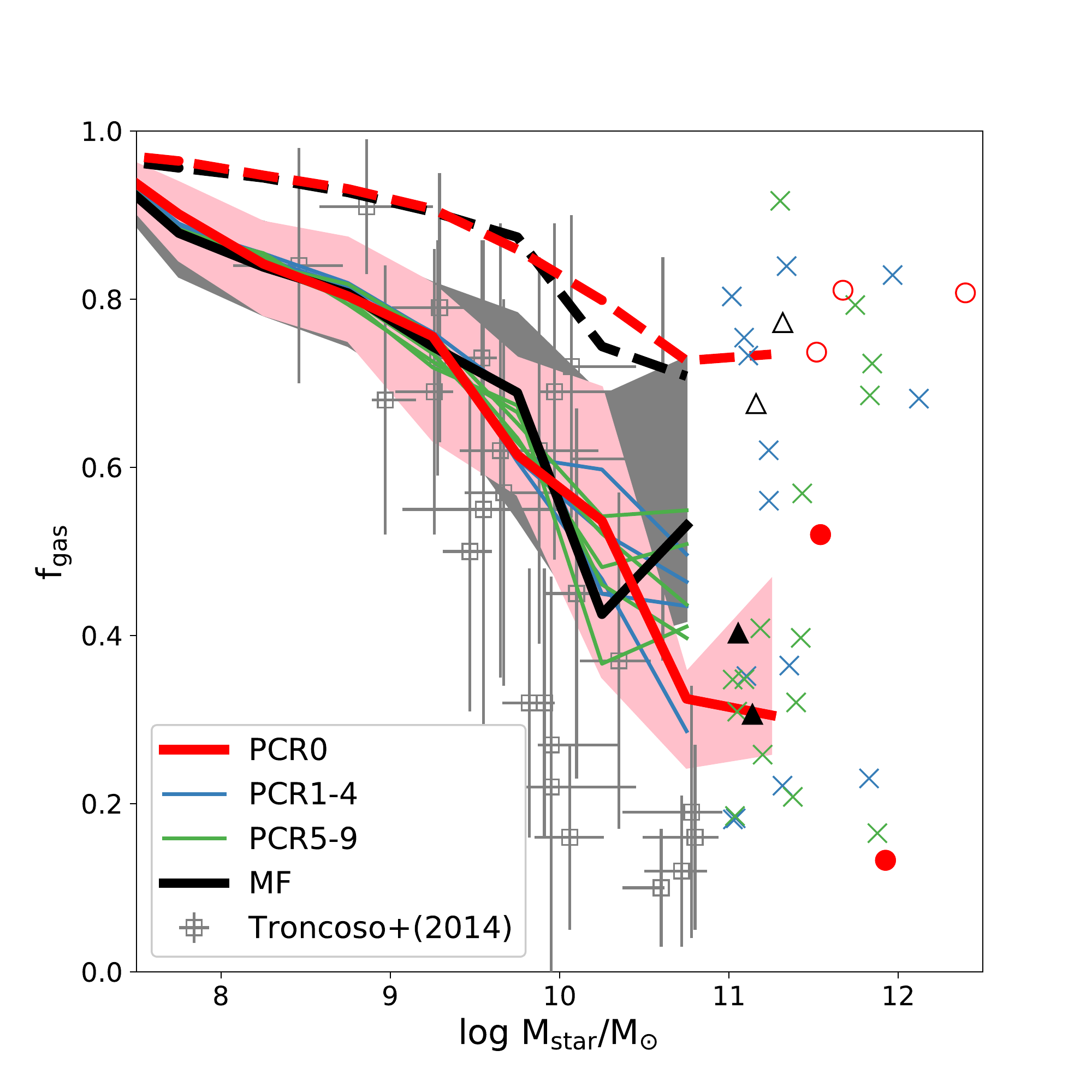}
	\end{center}
	\caption{	
	Gas mass to total baryon mass (gas + stars) fraction $f_{\rm{gas}}$  as a function of stellar mass at $z=3$. The pink and gray shades show the quartiles (25 - 75 percent) in each bin in PCR0 and MF runs. Different lines and symbols represent different runs, same as in figures \ref{fig:smhm}. 
	The red thick dashed line and open circles show the case using all gas and stars in haloes of PCR0. 
	The black thick dashed line and open triangles are the cases using all gas and stars in haloes of the MF run. 
	Gray open squares with error bars show the observed gas fractions in \citet{Troncoso14}. 
		 }
	\label{fig:fgas}
\end{figure}

\begin{figure}
	\begin{center}
		\includegraphics[width=9cm]{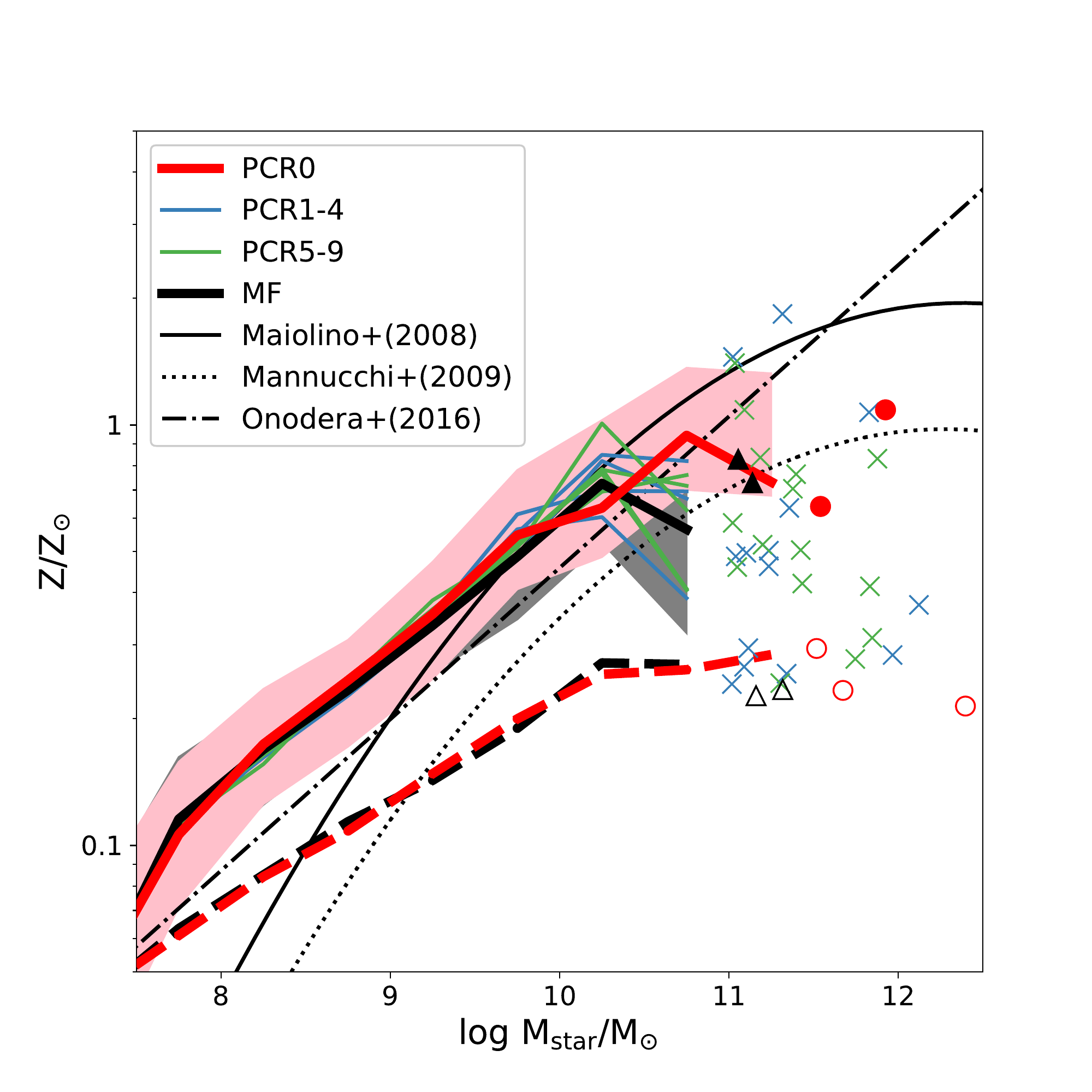}
	\end{center}
	\caption{	
	Gas metallicity as a function of stellar mass at $z=3$. Each line and symbol represent each run as in figure \ref{fig:fgas}. 	
	The metallicity and stellar mass are estimated within $2 \times r_{\rm 0.5}$ of the most massive member galaxies. The pink and gray shades show the quartiles (25 - 75 percent) in each bin in PCR0 and MF runs. 
	Black solid, dotted, dot-dashed lines represent the relations of observed galaxies derived in \citet{Maiolino08}, \citet{Mannucci09}, and \citet{Onodera16}. 
		 }
	\label{fig:metallicity}
\end{figure}

%===============================================
\subsection{Massive black holes in PC regions}

\begin{figure}
	\begin{center}
		\includegraphics[width=9cm]{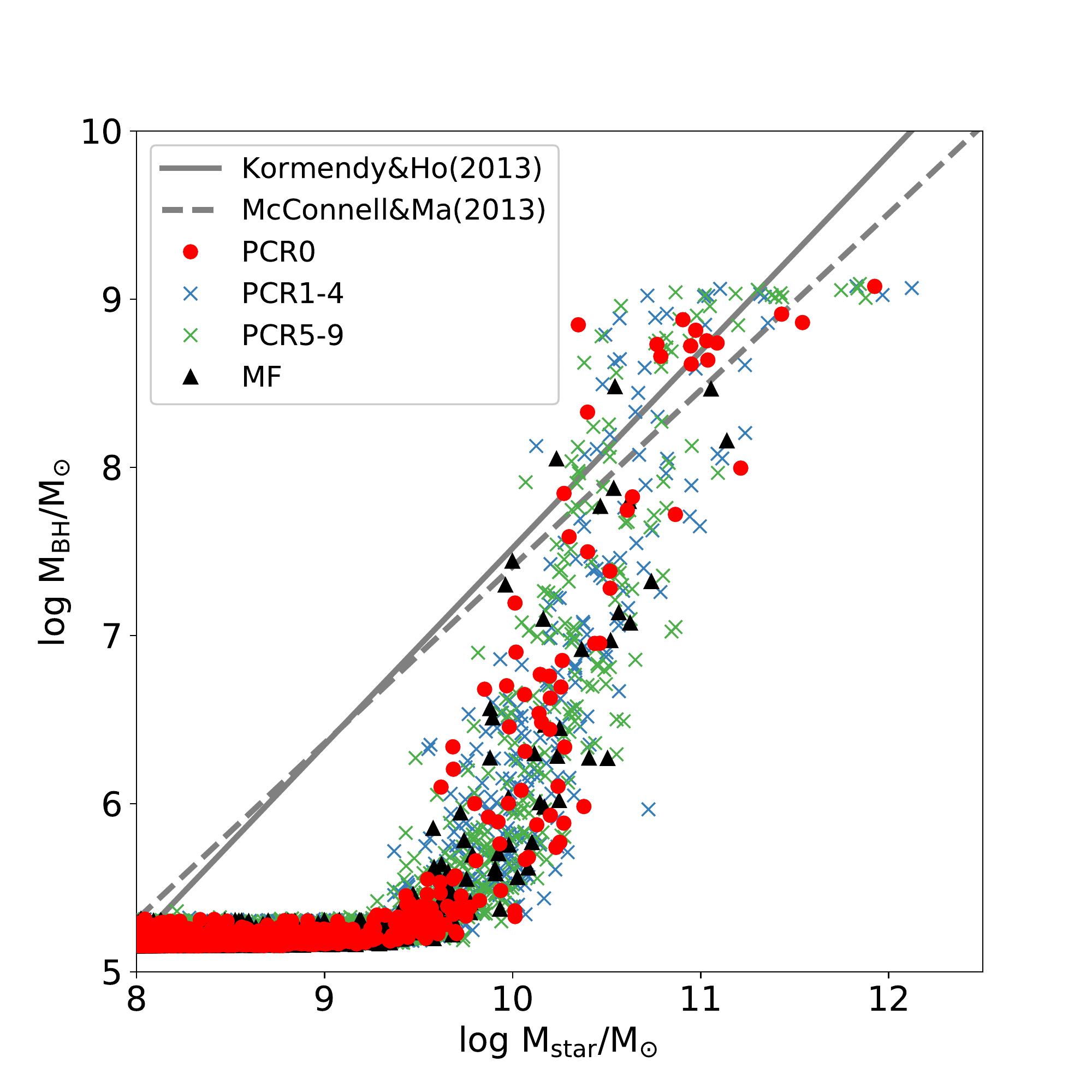}
	\end{center}
	\caption{		
		 Masses of the most massive black holes in each galaxy as a function of stellar mass at $z=3$. 
		 The gray solid and dashed lines represent the observed relations in local galaxies from \citet{Kormendy13} and \citet{McConnell13}, respectively.  
		 }
	\label{fig:bhmass}
\end{figure}

Massive BHs at galactic centres are ubiquitous in the local Universe \citep{Kormendy13}. 
The black hole mass is tightly correlated with the bulge mass of galaxies via $\Mbh \sim 2 \times 10^{-3}~\Mstar$ \citep[e.g.,][]{Marconi03}. While this correlation has been well established at low redshifts, it is still unclear how it looks at high redshift due to the limited number of observed massive black holes. 
Figure~\ref{fig:bhmass} shows the BH mass as a function of stellar mass. 
BHs grow slowly at $\Mstar \lesssim 10^{10}~\Msun$ and then do rapidly as the galaxies become more massive. 
As suggested by \citet{Dubois16},  SN feedback evacuates gas around a BH and suppresses the gas accretion onto it. 
Once the halo mass exceeds $\sim 10^{11-12}~\Msun$, the deep gravitational potential well associated with the halo keeps the gas confined at the galactic center against SN feedback. Therefore, the gas disc around the BH can become massive enough and allow  gas inflow to the galactic centre via clump formation and bar instability \citep{Shlosman89, Shlosman93}.
During this phase, the BHs grow at the Eddington limit and the BH mass increase as $\Mbh \propto {\rm exp} (t/t_{\rm Sal})$, 
where $t_{\rm Sal}$ is Salpeter time scale, $t_{\rm Sal} = \frac{f_{\rm r}\sigma_{\rm T}c}{4 \pi G m_{\rm p}}\sim 45~\rm Myr$. 
The energy injection rate is estimated as
\begin{equation}
\dot{E}_{\rm BH, feed} = 5.0 \times 10^{11}~{L_{\odot}}~\left(\frac{f_{\rm Edd}}{1.0} \right)
\left( \frac{f_{\rm e}}{0.15} \right) \left(\frac{\Mbh}{10^{8}~\Msun} \right). 
\end{equation}

Given that the gas accretion continues for a Salpeter time and a part of thermal energy is converted into kinetic one, the total kinetic energy is 
\begin{equation}
\begin{split}
E_{\rm kin} \sim 2.7 \times 10^{59}~{\rm erg} \left( \frac{f_{\rm conv}}{0.1}\right) \left( \frac{\Delta t}{45~\rm Myr}\right) \\
\times \left(\frac{f_{\rm Edd}}{1.0} \right)
\left( \frac{f_{\rm e}}{0.15} \right) \left(\frac{\Mbh}{10^{8}~\Msun} \right), 
\end{split}
\end{equation}
where $f_{\rm conv}$ is the conversion factor from thermal energy to the kinetic 
and $\Delta t$ as the accretion time scale of gas. 
On the other hand, the gravitational binding energy of the gas in a halo with $\Mh \sim 10^{13}~\Msun$ is estimated by
\begin{equation}
E_{\rm grav} \sim 1.1 \times 10^{59}~{\rm erg}~\left( \frac{\Mh}{10^{13}~\Msun}\right)^{2} \left( \frac{\xi_{\rm M}}{0.1}\right)^{2}
\left( \frac{\xi_{\rm gas}}{0.1}\right) \left( \frac{1+z}{4} \right),
\end{equation}
where 
$\xi_{\rm M}$ is the fraction of total matter mass within the star-forming region  (e.g., $\lambda \times R_{\rm vir}$ where $\lambda$ is the halo spin parameter \citep{Mo02}) to the total halo mass and
$\xi_{\rm gas}$ is the fraction of total gas mass to the total matter mass within the star-forming region. 
Therefore BH feedback can evacuate the gas from the star-forming region and suppress star formation although it does not continue for a long time due to the self-regulation of BH growth. 
Then, as the halo grows, galaxies can confine the gas and form stars (see also Figure~\ref{fig:mainseq}), while the growth of BHs is not so efficient due to the high-relative gas motion and low-gas density.
Some BHs reach $\sim 10^{9}~\Msun$ as their host galaxy mass increases. 
During this phase,  BH growth stalls  due to the powerful quasar and radio mode feedback while the stellar mass increases, ultimately leading to  massive galaxies with $\Mstar \gtrsim 10^{11}~\Msun$ having SMBHs with  masses as expected from the local relation.

The growth histories of BHs sensitively depend on the gas accretion, the feedback and the seeding models. Our simulations use the AGN models similar to EAGLE project, although there are some differences as e.g., including the radio mode feedback and depositing the feedback energy into a nearest gas particle. Thus, the trend of the BH growth is similar to the results in \citet{Schaye15} and \citet{Rosas-Guevara16}. As stated above, the early growth of BHs is suppressed due to the SN feedback \citep[see also,][]{Habouzit17, McAlpine17}. This was also reported in FIRE simulations \citep{Angles-Alcazar17}.  On the other hand, in IllustrisTNG, a more massive BH with $8 \times 10^{5}~\Msun/h$ is seeded in a halo with the mass of $5 \times 10^{10}~\Msun/h$. Also, the relative velocity between a BH and gas is ignored in the estimate of the accretion rate. Consequently, BH masses even in galaxies with  $\Mstar \lesssim 10^{10}~\Msun$ obey the observed relation of local galaxies even at high-redshifts \citep{Weinberger18}. Thus, the early growth of BHs with $\Mbh \sim 10^{5} - 10^{7}~\Msun$ can be sensitive to the sub-grid models \citep{Pillepich18, Li20b}. 
Even recent observations have suffered from investigating co-evolution of BHs of $\Mbh \lesssim 10^{8}~\Msun$ in high-redshift galaxies \citep{Izumi19}. Future observations with higher sensitivity and angular resolution may show the stellar mass of BH host galaxies and constrain the BH models in simulations. 

%===============================================

%===============================================

\subsection{Infrared properties}
\begin{figure}
	\begin{center}
		\includegraphics[width=8cm]{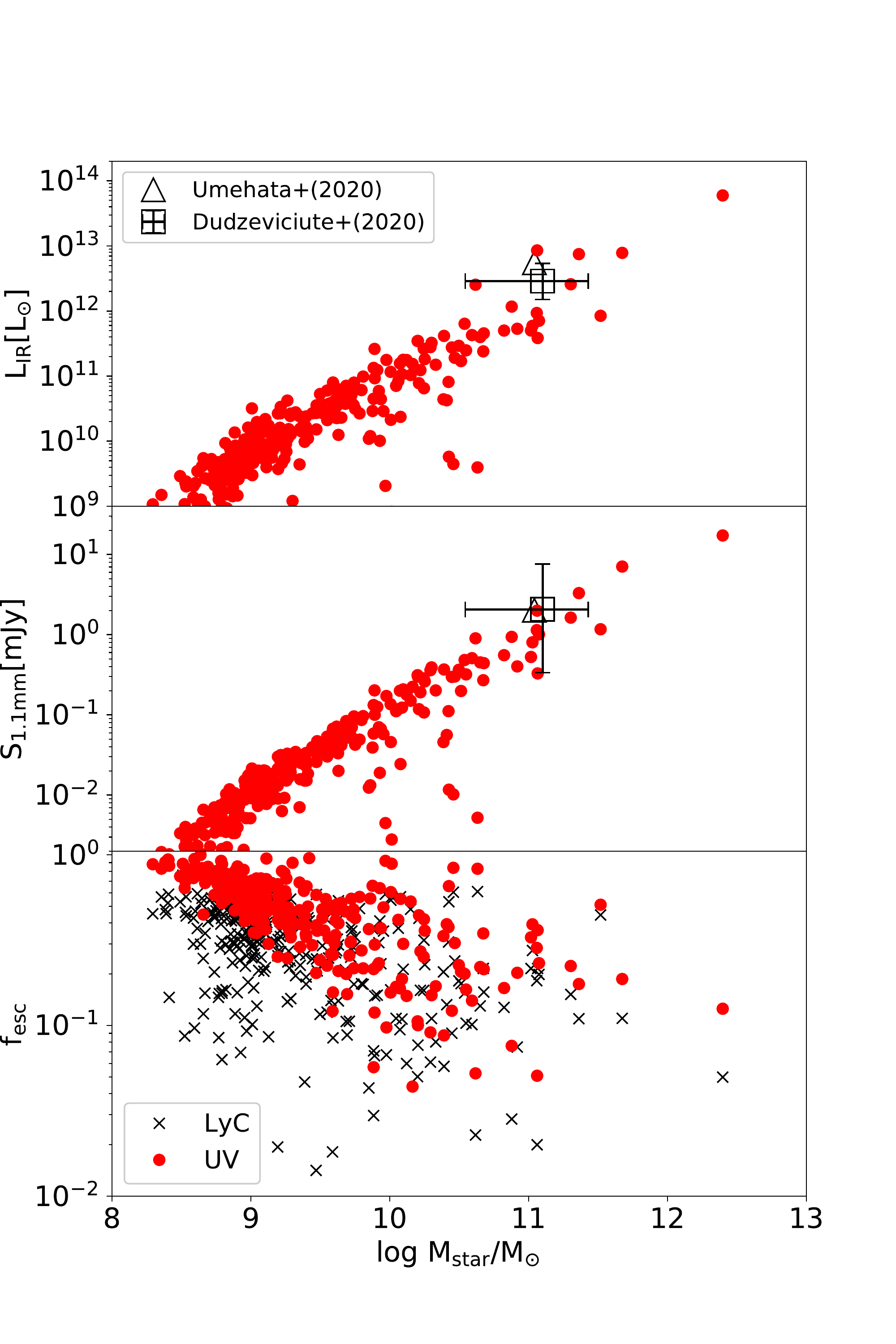}
	\end{center}
	\caption{		
		 Results of radiative transfer simulations of the 300 most massive galaxies
		 in PCR0 at z = 3. Top panel: Bolometric infrared luminosities as a function 
		 of total stellar mass in haloes. 
		 Middle panel: Submillimeter fluxes at $1.1~\rm mm$ in the observed frame. 
		 Open triangles represent a submillimeter galaxy at $z=4$ \citep{Umehata20}. Open squares with error bars show 707 submillimeter galaxies at $z = 1.8 - 3.4$ \citep{Dudzeviciute20} 
		 Lower panel: Escape fractions of UV (filled circles) and Lyman continuum (crosses) photons. 
		 }
	\label{fig:fesc}
\end{figure}

In order to investigate the observational signatures of the galaxies in the PC regions, 
we carry out radiative transfer simulations in post-processing.  
Figure~\ref{fig:fesc} presents IR luminosities ($\Lir$), fluxes at $1.1~\rm mm$ in the observed frame ($\Ssub$)
and escape fractions of UV and Lyman continuum photons. 
The escape fraction is estimated by counting the number of photon packets escaped from the calculation box ($\sim$ virial radius).
Here we estimate the radiative properties of the 300 most massive haloes in the PCR0 run. 
 $\Lir$ and $\Ssub$ increase with stellar mass. 
% The most massive halo has $\Lir = 6.0 \times 10^{13}~L_{\odot}$ and $\Ssub = 29.4~\rm mJy$. 
The most massive halo has $\Lir = 6.0 \times 10^{13}~L_{\odot}$ and $\Ssub = 17.3~\rm mJy$ ($S_{\rm 850\mu m}=30.8~\rm mJy$). 
 
 As the galaxy mass increases, dusty gas accumulates in star-forming regions and absorbs UV radiation efficiently. Therefore, $\fesc$ decreases as the stellar mass increases \citep[see also,][]{Vijayan21}. 
 At $\Mstar \gtrsim 10^{11}~\Msun$, $\fesc$ become smaller than $\sim 0.2$.  
Recently, \citet{Wang19} indicated that the fraction of dust-obscured galaxies becomes larger than UV bright galaxies (LBGs) at $\Mstar \gtrsim 10^{10.5}~\Msun$. Our results are consistent with their results. 
Due to the mass dependence of $\fesc$,  $\Lir$ and $\Ssub$  increase more steeply than the relation between SFR and $\Mstar$ in figure~\ref{fig:mainseq}. 
\citet{Umehata20} estimated the stellar mass and submillimeter flux of an SMG at $z=4.0$. In addition, \citet{Dudzeviciute20} successfully derived the physical properties of 707 SMGs at $z = 1.8 - 3.4$. 
Our modeled galaxies with similar stellar masses match those observations well. 
Recently, \citet{McAlpine19} investigated the properties of galaxies with the sub-mm flux at $850~\rm \mu m$ larger than $1~\rm mJy$ and showed that the median value of the stellar mass was $6.7 \times 10^{10}~\Msun$ at $z=2.8$, which is similar to our results.
\citet{Yajima15c} also showed the formation of dusty starburst galaxies at $z \gtrsim 6$. In \citet{Yajima15c}, we showed results for a massive galaxy with  $\Mstar = 8.4 \times 10^{10}~\Msun$ and $\Lir = 3.7 \times 10^{12}~L_{\odot}$ at $z = 6.3$,  
which are similar to our current results. In this paper, we have expanded the mass and redshift range, as well as added new sub-grid models.
Note that, if a halo contains multiple massive stellar components like, e.g., a major merger process, the submillimeter flux is likely to depend on the aperture size. 
Also, the submillimeter flux can be changed by lensing effects or blending with unassociated sources along the line of sight \citep{Hayward13}.
We will investigate such an aperture dependency of the submillimeter flux considering the quite different beam sizes of the current telescopes as ALMA and SCUBA2.

At $\Mstar \sim 10^{10}-10^{11}~\Msun$, there is a large dispersion in $\fesc$. 
Some galaxies have very high $\fesc$ of $> 0.5$ likely due to the galactic outflows. 
Therefore these galaxies are faint at sub-millimeter wavelengths with $\Ssub \lesssim 10^{-2}~\rm mJy$. 
This suggests that the population of galaxies in this mass range is not homologous. 
\citet{Arata19} showed that  SN feedback induces  galactic outflows and quenching of star formation
and the radiative properties rapidly changed due to this \citep[see also,][]{Yajima17c}. 
Also, \citet{Katsianis17} showed that luminosity functions (or SFR functions) sensitively depend on the SN/BH feedback models.
We will investigate the radiative properties and the origin of the observed diversity by using a larger galaxy sample in a subsequent paper. 

Note that, in the current simulations the multi-phase  ISM can not be resolved well due to the limited resolution. Therefore, {$\rm ART^{2}$} assumes a sub-grid model consisting of a two-phase ISM with cold gas clumps in a warm medium. In this case, the escape fraction can differ from the single-phase ISM model because some photons travel without interaction with the cold gas clump \citep{Yajima15c}. We will investigate the impacts of the ISM model on the radiative properties in future.  
However, since $\fesc$ of some massive galaxies is lower than $\sim 0.2$, their submillimeter fluxes do not change significantly even if $\fesc$ decreases furthermore.

%===============================================
%===============================================

\subsection{Baryon contents in galaxy clusters at $z \lesssim 1$}

To compare with the observations of local galaxy clusters, we carry out three additional simulations: 
PCR5($z_{\rm end} = 1.0$), PCR2low ($z_{\rm end}=0.0$, $m_{\rm gas}=2.3 \times 10^{7}~\Msun$ and $m_{\rm DM}=1.3 \times 10^{8}~\Msun$) and PCR6low ($z_{\rm end}=0.0$, $m_{\rm gas}=2.3 \times 10^{7}~\Msun$ and $m_{\rm DM}=1.3 \times 10^{8}~\Msun$). The zoom-in volume of original PCR runs can follow the growth of main progenitor haloes down to $z \sim 1$. Whereas, at $z \lesssim 1$, the masses of the most massive haloes differ from the results of the $N$-body simulations with a zoom-in volume of $(57.1~\rm cMpc)^{3}$. This indicates that the zoom-in regions do not cover all building blocks of the galaxy clusters at $z \lesssim 1$. 
Therefore, we recreate the initial conditions of PCR2 and PCR6 with a zoom-in volume of $(42.9~\rm cMpc)^{3}$ and perform the simulations down to $z=0$, although the mass resolutions of the SPH and dark matter particles are lower by a factor of 8. We confirm that the masses of the most massive haloes at $z_{\rm end}$ in PCR5, PCR2low, and PCR6low match the ones in the $N$-body simulations with the larger volume.

\begin{figure}
	\begin{center}
		\includegraphics[width=8cm]{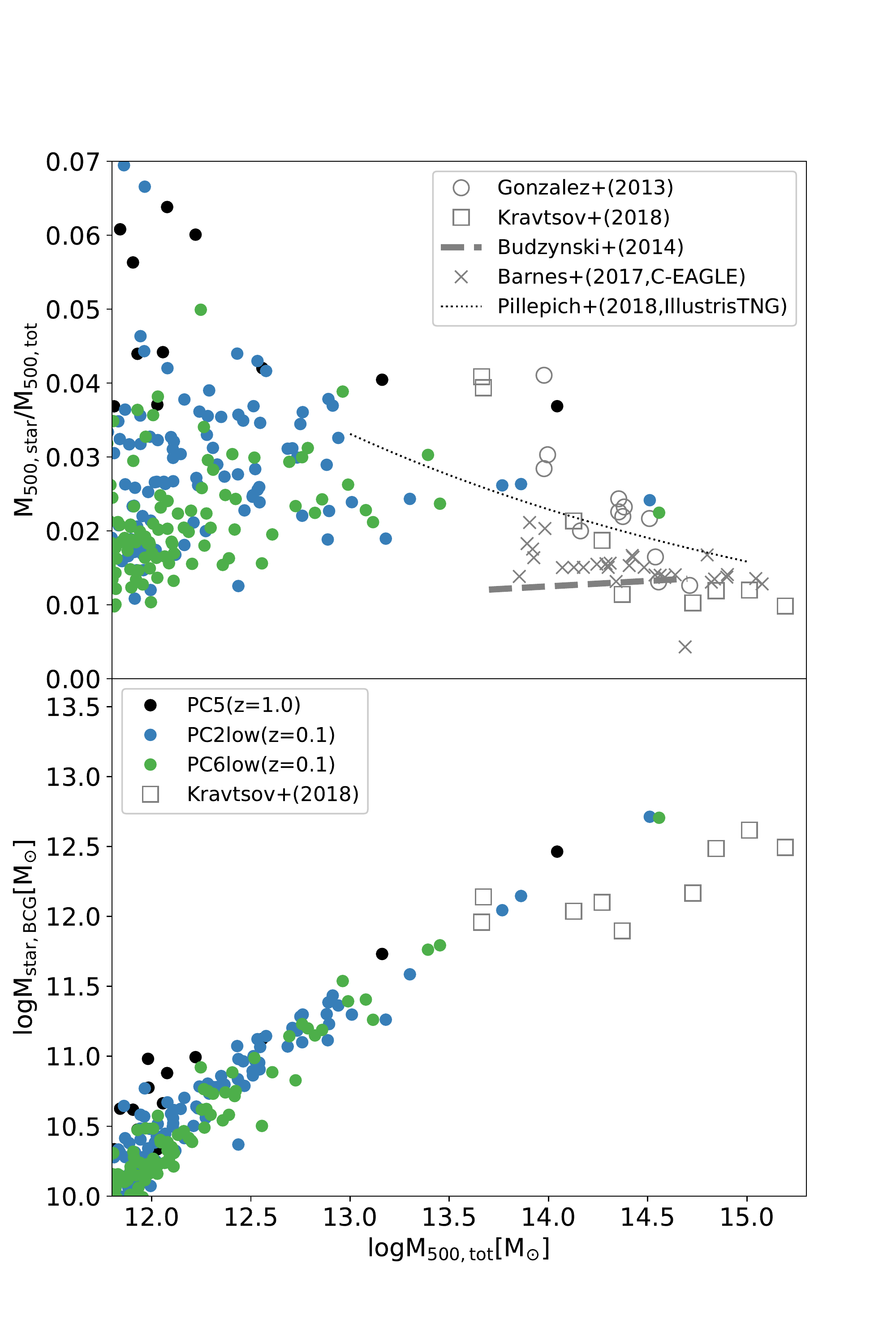}
	\end{center}
	\caption{		
		 Upper panel: Stellar mass fraction to halo mass. 
		 $M_{\rm 500, tot}$ and $M_{\rm 500, star}$ are total matter and stellar masses within the radius where the matter overdensity becomes 500. 
		 Filled circles are the simulation results: PCR5 at $z=1.0$ (black), 
		 PCR2low at $z=0.1$ (blue), and PCR6low at $z=0.1$ (green), 
		 where PCR2low and PCR6low runs use the larger zoom-in volume of $(57.1~\rm cMpc)^{3}$ 
		 than the original PCR2 and PCR6, and the mass resolutions of SPH and dark matter are 8 times lower. 
		 Doted line and crosses show the simulation results of IllustrisTNG \citep{Pillepich18b} and C-EAGLE \citep{Barnes17}, respectively.  Open circles, open squares and dashed line represent the observations of local galaxy clusters by \citet{Gonzalez13}, \citet{Kravtsov18} and \citet{Budzynski14}. 
		 Lower panel: Stellar masses of BCGs as a function of halo mass. The stellar masses are estimated within $2 \times r_{\rm 0.5}$ of the most massive member galaxies.
		 Open squares show the observations \citep{Kravtsov18}. 
		 }
	\label{fig:mstar_lowz}
\end{figure}

Figure~\ref{fig:mstar_lowz} presents the fractions of total stellar masses to total halo ones. 
Our simulations show that the stellar mass fractions of haloes with $\Mh \gtrsim 10^{13}~\Msun$ 
range $\sim 0.02 - 0.04$, and reproduce the observations \citep{Gonzalez13, Budzynski14,  Kravtsov18}, although some observed clusters have  low fractions of $\sim 0.01$. 
Also, the stellar mass fractions are similar to the results of 
the IllustrisTNG project \citep{Pillepich18}. On the other hand, our results are somewhat higher than those of the C-EAGLE project \citep{Bahe17, Barnes17}. 
This suggests that the total stellar mass in a halo is sensitive to the sub-grid models, the numerical resolution and the scheme. 
Lower panel shows the stellar masses of bright cluster galaxies in haloes. 
We estimate the stellar masses within $2 \times r_{\rm 0.5}$ of the most massive member galaxies. The stellar mass monotonically increase with the halo mass, which can be fit with a power-law function \citep{Pillepich18}. IllustrisTNG reports stellar masses of BCGs larger than in  observations \citep{Gonzalez13, Kravtsov18}. Our simulations also show a similar trend. The stellar masses of BCGs in host haloes with $M_{\rm 500, tot} > 10^{14}~\Msun$ are larger than those in observations by a factor of few. 
We find that the BCGs in PCR2low and PCR6low obtain half of their stellar masses at $z < 1.3$ and $1.1$, respectively. Therefore, the star formation activity can be too high in the later phases of the PCs. 
Suppression mechanisms of star formation in galaxy clusters at low redshifts will be investigated in future work.

\begin{figure}
	\begin{center}
		\includegraphics[width=8cm]{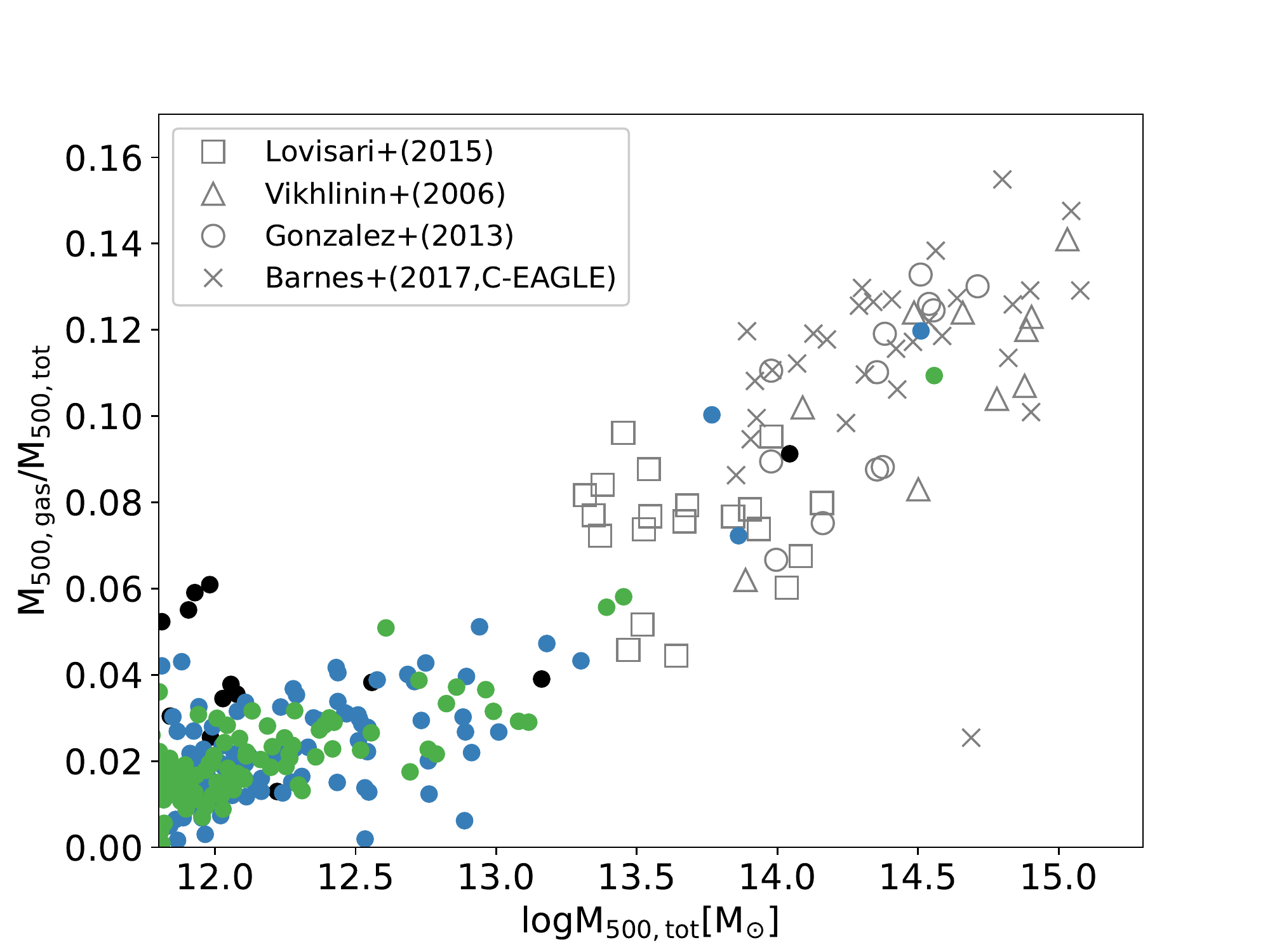}
	\end{center}
	\caption{		
		  Gas mass fraction to halo mass. Filled circles are the simulation results: PCR5 at $z=1.0$ (black), 
		 PCR2low at $z=0.1$ (blue), and PCR6low at $z=0.1$ (green) as in Figure~\ref{fig:mstar_lowz}.
		   $M_{\rm 500, gas}$ is total gas mass within the radius where the matter overdensity becomes 500. 
		 Open squares, triangles, and circles show the observations of local galaxy clusters by \citet{Lovisari15}, \citet{Vikhlinin06}, and \citet{Gonzalez13}. Filled circles are our simulation results and the same as Figure~\ref{fig:mstar_lowz}.
		 }
	\label{fig:mgas_lowz}
\end{figure}

The gas mass fractions in haloes are presented in Figure~\ref{fig:mgas_lowz}.
The gas mass fractions of our simulations increase with the halo mass
and show $\sim 0.1$ for massive haloes with the mass of $\sim 10^{14}~\Msun$
that nicely match the observations \citep{Vikhlinin06, Gonzalez13, Lovisari15}.
Note that, the observed physical properties within specific radii are estimated from surface brightness with the assumption of  spherically symmetric density profiles \citep{Lovisari15}. Therefore, if the gas and temperature structures of galaxy clusters differ from the spherical symmetry, the estimates changes somewhat.
In the case of the gas mass fractions, our results are somewhat lower than C-EAGLE. 
This indicates that our simulations more efficiently convert gas into stars or evacuate  gas from haloes.  
Note that, the number of massive haloes in our simulations is not enough to discuss the statistical nature of such trends.
The dependence of the sub-grid models or the simulation schemes on the baryon contents in local galaxy clusters  will be investigated further more in future work. 

%===============================================

%===============================================

\subsection{BCG runs}
\label{sec:BCG}
\begin{figure}
	\begin{center}
		\includegraphics[width=8cm]{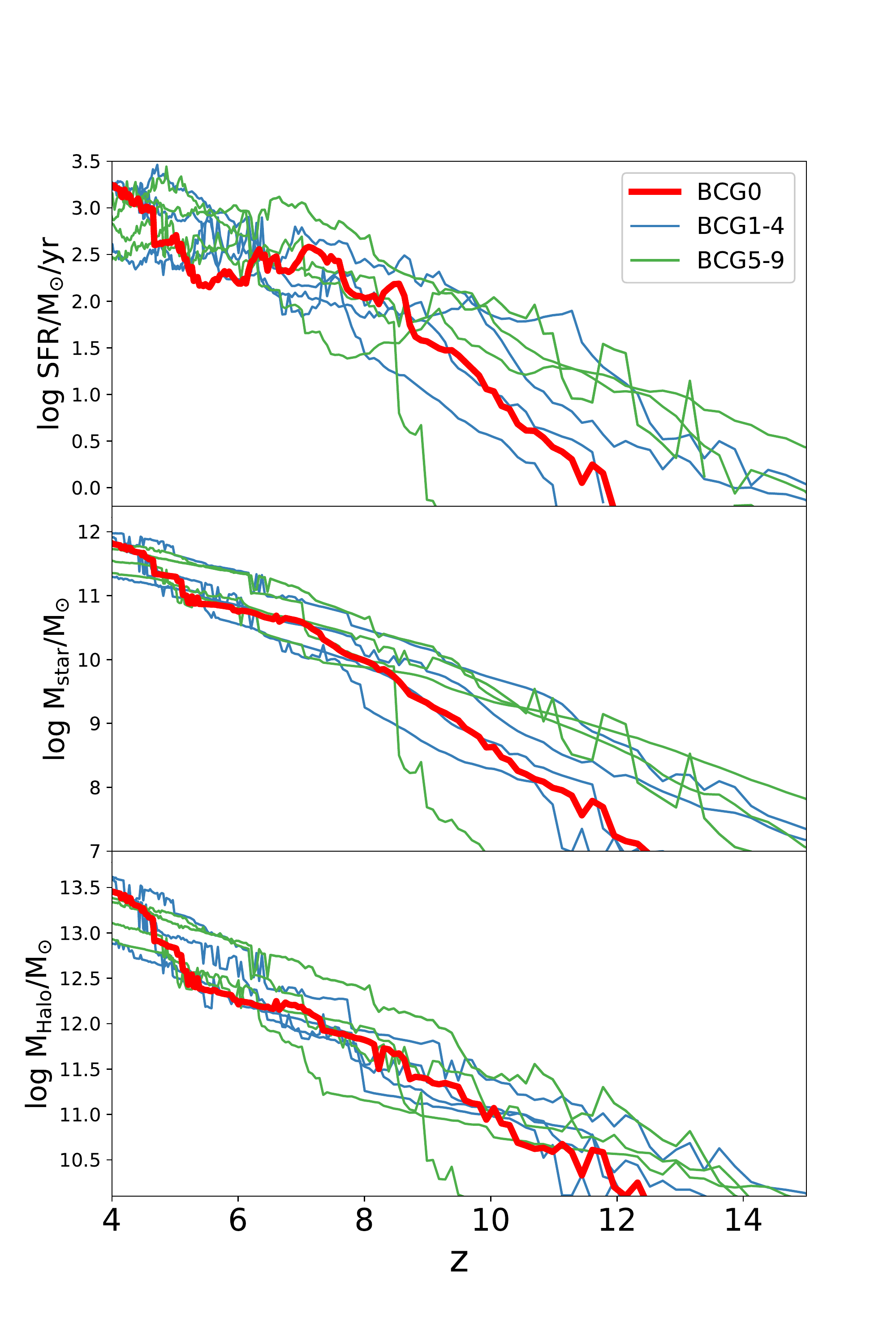}
	\end{center}
	\caption{		
		Redshift evolution of SFR, stellar mass and halo mass of BCG runs.  
		 }
	\label{fig:bpcg}
\end{figure}

We study the time evolution of the most massive haloes in the BCG runs. 
Here, we evaluate total quantities in a halo, e.g., SFR refers to the total SFR in a halo.
Figure~\ref{fig:bpcg} shows the star formation histories, stellar mass, and halo mass growth histories.
Here, we identify a main progenitor halo as the most massive halo with more than 30 percent overlapping dark matter particles  between the current and next snap shots. 
The halo masses of the BCGs exceed $\sim 10^{12}~\Msun$ even at $z \sim 7$
and reach $\Mh \sim 1- 3 \times 10^{13}~\Msun$ at $z \sim 4$. 
All BCGs host galaxies with $\Mstar \gtrsim 10^{11}~\Msun$ at $z \lesssim 6$. 

The SFR of BCG0 increases from $z \sim 10$ to $\sim 7$ significantly. 
This is because the halo can keep confining the gas against  SN feedback as the halo mass  becomes close 
to $\Mh \sim 10^{12}~\Msun$ (see also figure~\ref{fig:smhm}).
The SFR stays at  $\sim 100-300~\Msunyr$ at $z \sim 5-7$ while the halo grows slowly. 
At $z \lesssim 5$, the halo mass of BCG0 increases rapidly, resulting in a starburst with $\rm SFR \gtrsim 1000~\Msunyr$. 
Most  BCGs have high SFRs with $\gtrsim 100~\Msunyr$ even at $z = 6 - 8$, which is similar to  observed dusty starburst galaxies \citep[e.g.,][]{Walter18}. 
This suggests that observed dusty starburst galaxies form in protoclusters. 
BCG7 has the highest value of SFR at $z > 6$, which is $\rm SFR = 1254~\Msunyr$ at $z = 6.4$.  
This is similar to the bright SMG at $z=6.3$, HFLS3 \citep{Riechers13, Cooray14}. 

Recent observations indicated that passive galaxies form after the starburst phase \citep{Glazebrook17,  Mawatari20}. 
However, all BCGs in our simulations keep high SFRs at $z \lesssim 6$. In our simulations, even if  SN or BH feedback suppresses star formation for a while, dark matter and gas keep accreting on the haloes and avoid quenching of star formation for a long time ($\gtrsim 1~\rm Gyr$). 
Our result thus suggests that  it may require a rare situation where the growth rate of a halo is quite small for a long time. 
We will investigate such setups using a larger sample in future work. 

\begin{figure}
	\begin{center}
		\includegraphics[width=8cm]{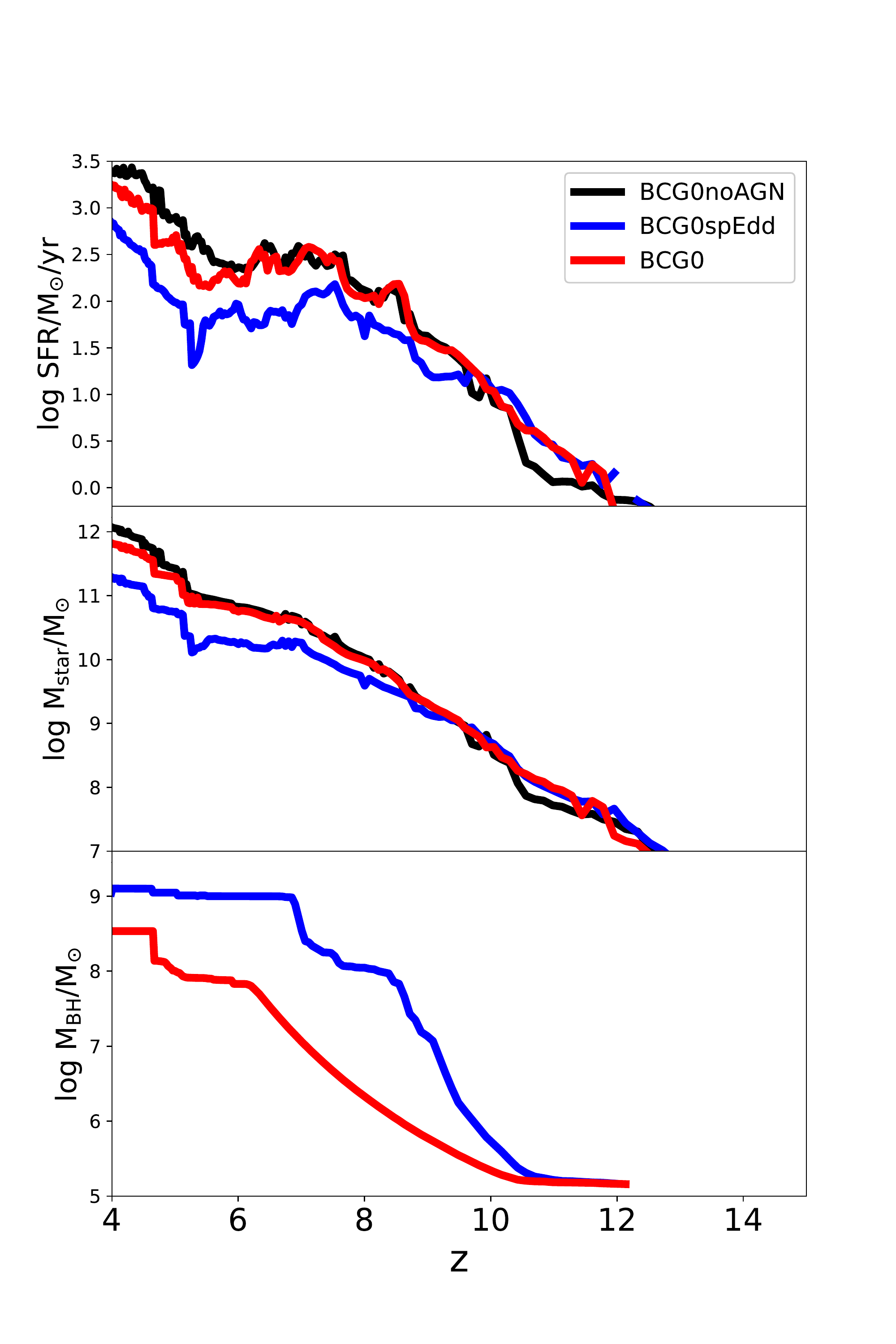}
	\end{center}
	\caption{		
		Same as figure~\ref{fig:bpcg}, but comparing BCG0, BCG0noAGN and BCG0spEdd. 
		 }
	\label{fig:bpcg_model}
\end{figure}

In most  BCGs, SMBHs with $\Mbh > 10^{8}$ form at $z \lesssim 6$. 
Therefore, BH feedback can play a role in regulating star formation and shaping the gas structure. 
Figure~\ref{fig:bpcg_model} shows the star formation histories, the growth histories of stellar and BH mass 
of BCG0, BCG0noAGN, and BCG0spEdd. 
Given that the upper limit of the Eddington ratio is set to $5$ (BCG0spEdd), 
the BH mass  rapidly increase from $\sim 10^{5}$ to $\sim 10^{8}~\Msun$ at $z = 8 - 10$. 
Then it achieves $ 10^{9}~\Msun$ at $z  = 6.5$ after the stalling phase. 
Due to the self-regulation via the quasar and radio mode feedback processes, 
the growth of the BH becomes slow. Finally, the mass of the BH in BCG0spEdd is $1.3 \times 10^{9}~\Msun$ at z=4.0. 
On the other hand, BCG0 hosts the BH with $\Mbh = 6.7 \times 10^{7}~\Msun$ even at $z =6.0$. 
The growth rate of the BH mass becomes small at $z = 4.8 - 6.2$ when the halo growth is slow. 
At $z < 5$, the BH mass increases via the merger of BHs and achieves $\Mbh = 3.4 \times 10^{8}~\Msun$. 
Therefore, we suggest that the growth history of a BH depends on the upper limit of the accretion rate.  
The upper limit is likely to be determined by unresolved small-scale structure, i.e., the gas distribution, angular momentum, and anisotropy of the radiation from an accretion disk. 
If the gas structure and the flux from an accretion disk are isotropic, the accretion rate should not exceed the Eddington limit \citep[but see,][]{Inayoshi16}. On the other hand, given that the anisotropies of gas and radiation, the accretion rate can be estimated by the Bondi-Hoyle-Littleton model and be larger than the Eddington limit \citep[e.g.,][]{Sugimura17}. 

BH feedback suppresses the star formation as shown in the upper panel. 
In the case of BCG0, the SFR becomes smaller than BCGnoAGN by a factor of 1-3 at $z \lesssim 6$.  
At $z  > 6$, the difference of the SFRs is quite small although the BH grows almost at the Eddington limit at $z \sim 6 - 10$. This suggests that the injected thermal energy is lost efficiently by radiative cooling before it induces a large-scale galactic outflow. 
In the case of BCG0spEdd, the reduction rate of the SFR is much larger, it is lower than BCGnoAGN by an order of unity at $z \lesssim 7$. As a result, $\Mstar$ of BCG0spEdd is lower than BCG0noAGN by a factor of 3.5 at $z=4.0$, while there is no large difference between BCG0 and BCG0noAGN. 

\begin{figure}
	\begin{center}
		\includegraphics[width=8cm]{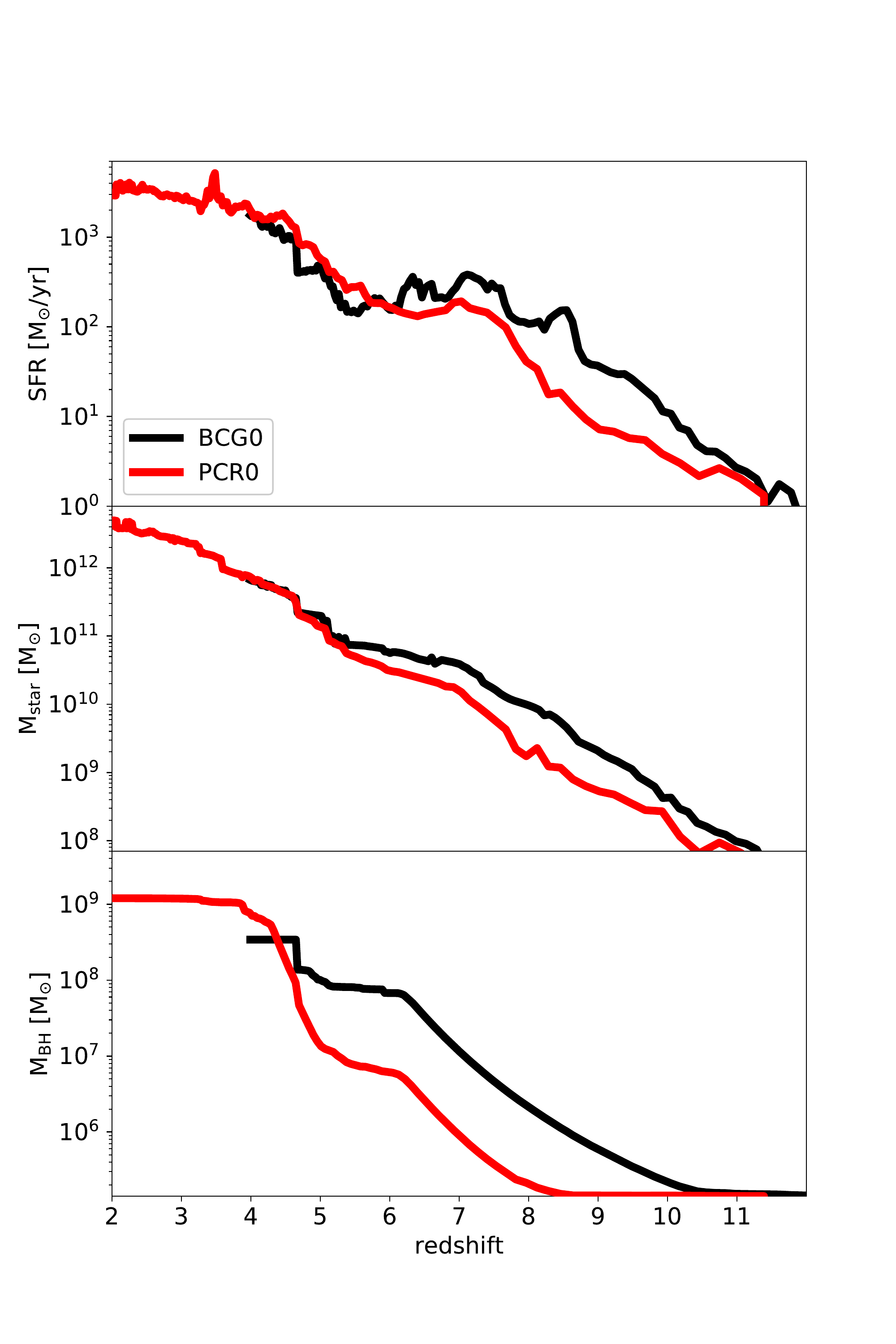}
	\end{center}
	\caption{	
	Redshift evolutions of star formation rate (top), stellar mass (middle), and black hole mass (bottom) in main progenitors of the most massive haloes at the final redshifts.
	Black and red lines represent BCG0  and PCR0 runs, respectively.   
		 }
	\label{fig:sfr_difrun}
\end{figure}

In addition, we compare the results of BCG0 with PCR0 as a resolution study. 
Figure~\ref{fig:sfr_difrun} represents the redshift evolution of SFRs, stellar masses, and BH masses in main progenitors.
We show the SFR of BCG0 is higher than that of PCR0 by a factor of few at $z > 6$. The difference becomes smaller at lower redshifts $z \lesssim 6$. 
The difference of the cumulative stellar mass is a factor of few at $z \sim 8$ and becomes quite small at $z \lesssim 5$. 
Therefore we suggest that the star formation history is not sensitive to the resolution at $z \lesssim 6$. 
In the case of BHs, the early growth sensitively depends on the resolution. The BH mass in PCR0 does not grow down to $z \sim 8$, while it starts to grow at $z \sim 10$ in BCG0. 
The mass dependence of the Bondi-Hoyle-Littleton model is $\propto \Mbh^{2}$. 
To keep high-gas accretion-rates close to the Eddington limit for seed mass BHs ($10^{5}~\Msun \; h^{-1}$), 
high-density gas with hydrogen number density of $\gtrsim 100~\rm cm^{-3}$ is needed at a galactic centre.
Therefore, the earlier growth of BHs indicates that BCG0 can resolve the high-density gas region at the galactic centre successfully.  
Note that, however, there is no significant difference in the BH mass at $z \sim 4$ between PCR0 and BCG0. 
Thus, the simulation results at $z \sim 3$, which we mainly focus on in this paper, is unlikely to be sensitive to the numerical resolution. 

%===============================================

%===============================================
%===============================================
%===============================================

\subsection{First runs}
Metal enrichment of the universe proceeds inhomogeneously \citep{Wise12a, Pallottini14, Hicks20}.
The overdensity regions are likely to be metal-enriched earlier than the mean-density field. 
Therefore, the transition from Pop III to Pop II stars occurs earlier. 
Here, we investigate the transition of the stellar population. 

\begin{figure}
	\begin{center}
		\includegraphics[width=8cm]{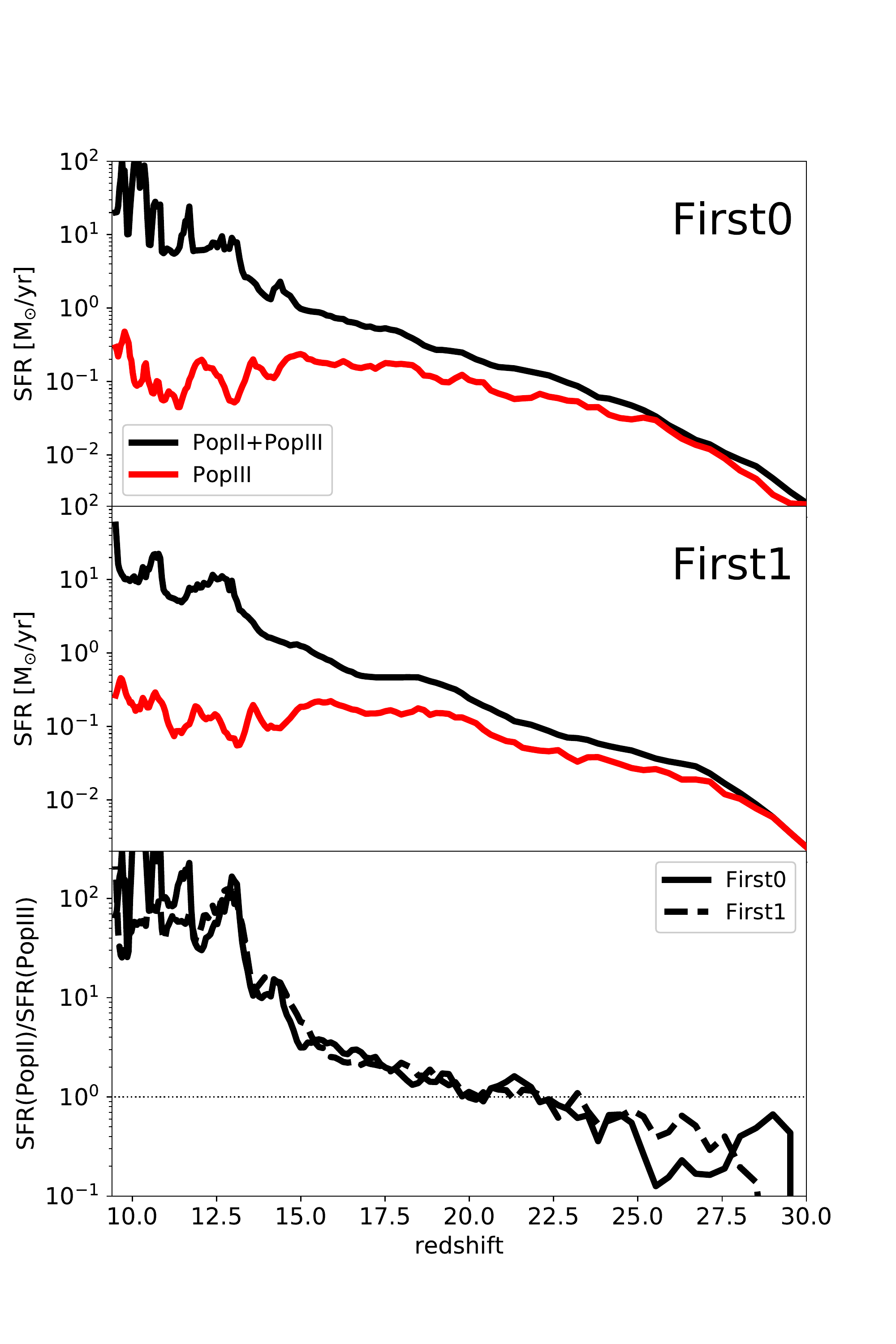}
	\end{center}
	\caption{	
	Star formation rate histories of First0 and First1 runs. 
	Bottom panel shows the ratio of star formation rates of Population II to Population III stars. 
		 }
	\label{fig:sfr_first}
\end{figure}

Figure~\ref{fig:sfr_first} presents the total star formation rates of Pop III and Pop II stars 
in the zoom-in regions. At $z \sim 30$, Pop III stars form gradually with a rate of $\sim 10^{-2}~\Msunyr$.
Due to the SNe of PopIII stars, the gas is metal-enriched, and Pop II stars start to form at $z \sim 25$.
The total SFRs of Pop III stars keep increasing up to $z \sim 15$. 
Then it decreases gradually at $z \sim 10-15$. 
On the other hand, the SFR of Pop II stars increases with time monotonically. 
The SFR of Pop II stars exceeds that of Pop III stars at $z \sim 20$. 
It is earlier than the mean density field, $z \sim 15$, as shown in Johnson et al. (2013). 
Also, as the SFR increases, mini-haloes with pristine gas are irradiated by strong LW radiation from star-forming galaxies, resulting in the suppression of the formation of Pop III stars. 
At $z \sim 10$, the SFR of Pop II stars become $\sim 100$ times higher than that of Pop III stars. 

Because of the rapid halo growth in the overdensity regions, 
the most massive haloes form stars actively with $\rm SFR \gtrsim 18~\Msunyr$ even at $z \gtrsim 10$. 
These galaxies can emit strong Ly$\alpha$, H$\alpha$ lines, and metal lines. 
Therefore they can be prime targets in future observations with ALMA, JWST, and other 30-m class telescopes (e.g., E-ELT, TMT, GMT). 

Also, wide-field near-infrared imaging surveys would be key to finding such rare overdense regions. Future missions with e.g. Euclid and the Roman space telescopes will be expected to search for such regions.

%===============================================
%===============================================

%----------------------------------------------------------------------
%
% Section 4:  Discussion
%
%----------------------------------------------------------------------

%\section{Discussion}

%in prep. 

%----------------------------------------------------------------------
%
% Section 5:  Summary
%
%----------------------------------------------------------------------

\section{Discussion \& Summary}

In this paper, we introduce a new simulation project FOREVER22: 
FORmation and EVolution of Extremely-overdense Regions motivated by SSA22.
In this project, we study galaxy evolution in protocluster (PC) regions using cosmological hydrodynamics simulations with zoom-in initial conditions. 
FOREVER22 consists of three types of runs with the different resolutions and  zoom-in volumes. 
Using these simulations, we study the statistical natures of galaxies in PCs, gas dynamics of individual galaxies, and feedback processes. 
We select 10 protocluster regions from a cosmological  
box of size of $L=714~\rm cMpc$. 

The main conclusions of this paper are the following  points: \\ \\
1) In the PC regions at $z = 3$, the most massive halo reaches a halo mass of $\Mh = 1.2 \times 10^{14}~\Msun$ and hosts a super-massive black hole (SMBH) with $\Mbh = 1.2 \times 10^{9}~\Msun$. 
BHs grow rapidly as the host stellar mass exceeds $\sim 10^{10}~\Msun$. Then, the growth of supermassive BHs is suppressed due to their feedback, while the host stellar mass continues to increase. BH masses in massive haloes follow the observed local BH mass and bulge mass relation \citep[e.g.,][]{Marconi03}.   
\\ \\
2)  More than five starburst galaxies with ${\rm SFR} \gtrsim 100~\Msunyr$ form in the massive haloes with $\Mh \gtrsim 10^{13}~\Msun$ at the core of a PC region at $z=3$.
The most massive halo has a cumulative SFR of $2679~\Msunyr$. These massive active galaxies are dust-obscured, resulting in the bright submillimeter flux densities of $\gtrsim 1~$ mJy at $1.1~ \rm mm$. 
\citet{Yajima15c} investigated  the formation of dusty massive galaxies with $\Mstar \sim 8.4 \times 10^{10}~\Msun$ at $z = 6.3$. The infrared luminosities of our modelled massive galaxies are quantitatively similar to the previous study. In this work, we have expanded the ranges of redshift and halo mass with new sub-grid models including massive BHs. 
\\ \\
3) The metal enrichment proceeds efficiently via type-II supernovae in the early Universe and the dominant stellar population  changes from Pop III to Pop II at $z \sim 20$. In the metal-enriched PC cores, the first galaxies with  ${\rm SFRs} \gtrsim 18~\Msunyr$ form at $z \sim 10$.\\\\

Thus, we suggest that  PCs can be the formation sites of bright submillimeter galaxies and SMBHs at $z \sim 3$. 
The clustering of dusty galaxies are similar to the one in observed protoclusters, e.g., SSA22 region. 
In addition, the bright first galaxies at $z \gtrsim 10$ can be prime targets for future observations by {\it James Webb Space Telescope. \\
}
 
 In this paper, we mainly present an overview of the properties of galaxies  
 in the PCs at $z=3$. Recently, \citet{Bouwens20} showed the contribution of bright SMGs to the cosmic SFR density over a wide redshift range \citep[see also,][]{Wang19}. They indicated that the contribution becomes much smaller than that of UV-selected galaxies at $z \gtrsim 4$. This is closely related to the redshift evolution of the star formation activity and dust distribution in massive haloes in PCs. 
 In addition, the cosmic reionization proceeds inhomogeneously, and the over-dense regions can form ionized bubbles earlier in in-side out fashion \citep[e.g,][]{Iliev12}. Therefore, the PCs can be the first triggers of  reionization, although the escape probability of ionizing photons decreases as the halo mass increases \citep[e.g.,][]{Yajima11, Yajima14c, Wise14, Paardekooper15, Trebitsch17, Ma20}. 
 We will investigate other statistical properties over a wide redshift range and the origin of the observed diversity of high-redshift galaxies in a future papers.  

In the PCs regions, the growth rate of halo mass is much faster than in the mean-density field, resulting in the formation of massive haloes with $\gtrsim 10^{13}~\Msun$ at $z \sim 3$.
However, at a specific stellar or halo mass, the star formation, the gas fraction, and the metallicity of the PCs are similar to those of the mean density field, although some galaxies in the PCs have lower SFRs and have more massive BHs. Therefore, we suggest that the external environmental effects (e.g., galaxy merger, tidal force) on the properties such as stellar mass and star formation rate are not significant.

The massive galaxies in the PC regions show normal star formation activity, lying along the observed star formation  main-sequence. In particular, the dispersion of sSFRs of massive galaxies is not significant.
On the other hand, recent observations indicated that some massive galaxies enter the passive phase even at $z \gtrsim 2$ \citep{Glazebrook17}. 
In the current simulations, although stellar and AGN feedback evacuate gas from star-forming regions, the circum-galactic medium or IGM filaments feed galaxies with gas, resulting in continuous star formation. Therefore, 
the early quenching of star formation is likely to depend on the feedback model. Stronger feedback can delay the refueling time-scale and may induce the formation of the passive galaxies. Due to the limited resolution, AGN feedback is modeled via a sub-grid model with free parameters, e.g., the thermal coupling factor. 
In addition, \citet{Yajima17c} showed that the higher amplitude factor in the star formation model induced large fluctuations in star formation history. 
We will investigate the quenching mechanism of massive galaxies at high-redshifts by changing these conditions in our future work.

%----------------------------------------------------------------------
%
% Acknowledge
%
%----------------------------------------------------------------------
\section*{Acknowledgments}

We wish to thank the anonymous referee for detailed comments and suggestions that improved this paper. 
We are grateful to Masayuki Umemura, Ken Ohsuga and Kazuyuki Sugimura for valuable discussion and comments. The numerical simulations were performed on the computer cluster, XC50 in NAOJ, and Trinity at Center for Computational Sciences in University of Tsukuba. This work is supported in part by MEXT/JSPS KAKENHI Grant Number 17H04827, 20H04724, 21H04489 (HY), 17H01111, 19H05810, 20H00180 (KN), 17H0481, 17KK0098, 19H00697 (YM), 20H01953 (HU), 20K14530 (MK), NAOJ ALMA Scientific Research Grant Numbers 2019-11A, Leading Initiative for Excellent Young Researchers, MEXT, Japan (HJH02007), Spanish Ministry of Science and Innovation (MICIU/FEDER)  RYC-2015-18078 and PGC2018-094975-C22 (CDV), and JST FOREST Program, Grant Number JPMJFR202Z (HY). 

\section*{Data availability}

The data underlying this article will be shared on reasonable request to the corresponding author.

%----------------------------------------------------------------------
%
% References
%
%----------------------------------------------------------------------
\bibliographystyle{mn}
%\bibliography{mn-jour,HY}

\appendix
\section{Convergence test}\label{sec:appendix}
\begin{figure}
	\begin{center}
		\includegraphics[width=8cm]{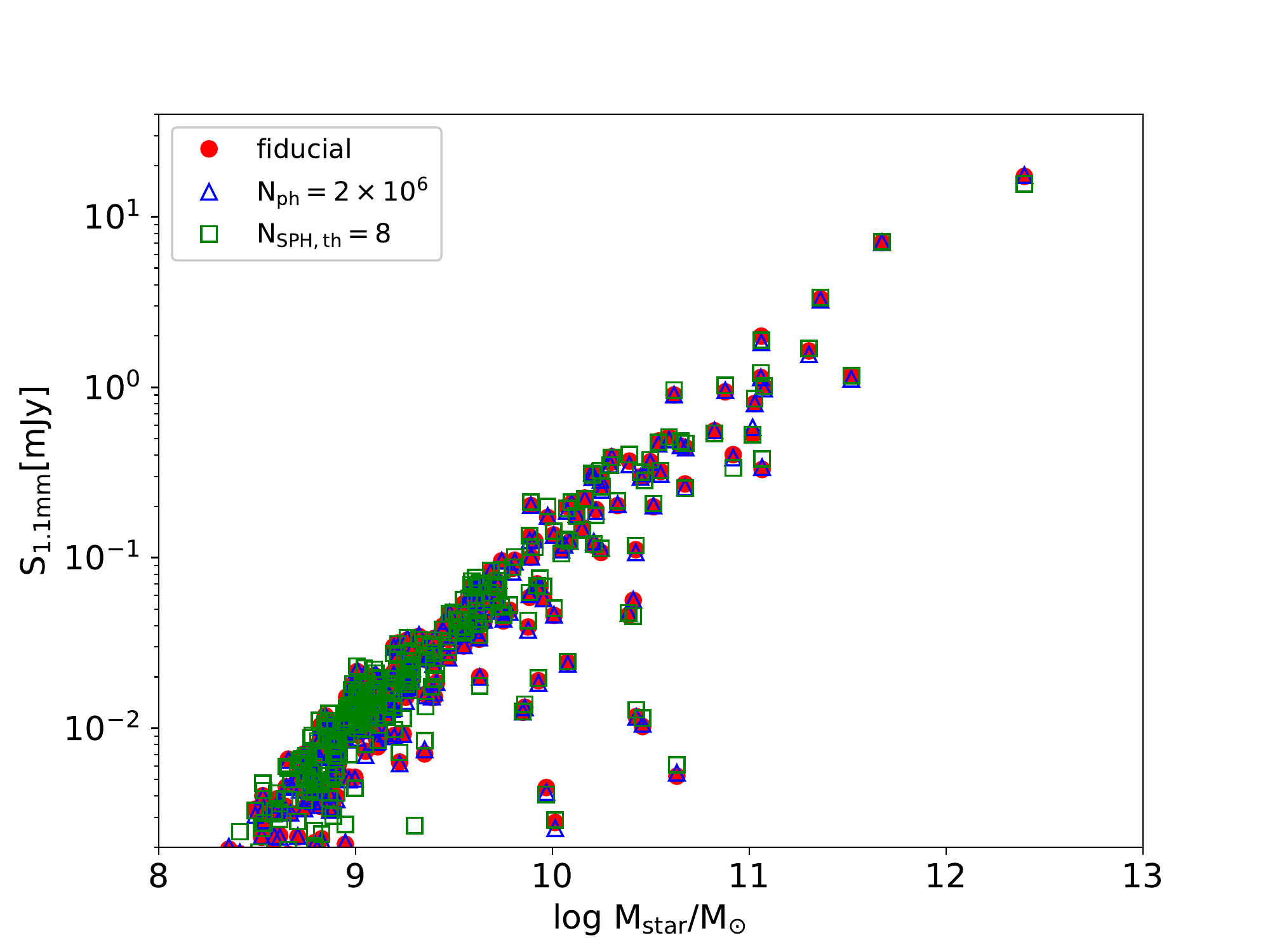}
	\end{center}
	\caption{		
		 Submillimeter fluxes at $1.1~\rm mm$ in the observed frame. Massive 300 haloes in PCR0 run are used. Red filled circles show the results of the fiducial runs. Open triangles and squares represent the cases with doubling photon packets and higher resolution of spatial grids.
		 }
	\label{fig:fsub_comp}
\end{figure}

Radiative transfer simulations based on Monte Carlo method, which is used in {\sc art2}, can involve a large dispersion of results due to the stochastic random sampling manner, if the number of photon packets is not large enough \citep{Iliev06, Yajima12a}. 
Therefore, we test the convergence to the number of photon packets. 
Figure~\ref{fig:fsub_comp} presents the submillimeter fluxes at $1.1~\rm mm$ in the observed frame.
We compare our fiducial simulations with  simulations using double as many photon packets ($2 \times 10^{6}$). We find that the results do not change significantly. The relative errors, $(S_{\rm 1.1mm}-S_{\rm 1.1mm}^{\rm fiducial})/S_{\rm 1.1mm}^{\rm fiducial}$, are mostly within $\sim 10$ percent. 

Besides the above, we investigate the dependency on the resolution of spatial grids. 
To perform the radiative transfer simulations, we construct the adaptive refinement grid structure from SPH particles. In making the structure, the resolution of the spatial grid is arbitrary. Our fiducial simulations set the number of base grid as $N_{\rm base}=4^{3}$ and make higher resolution grids if a cell around the grid contain more than $N_{\rm th}=16$ SPH particles. We additionally perform the simulations with $N_{\rm base}=8^{3}$ and $N_{\rm th}=8$.
As shown in Figure ~\ref{fig:fsub_comp}, the dependency on the spatial resolution is not significant. We find that the relative errors are mostly less than 20 percent. Thus, although the results can change somewhat depending on the number of photon packets and the grid resolution, the statistical properties of the modeled emergent fluxes from galaxies do not change. 

%===============================================
%===============================================

\label{lastpage}

\end{document}